\documentclass[a4paper,10pt,twoside]{article}
\usepackage{latexsym}
\usepackage{amsmath}
\usepackage{amssymb}
\usepackage{amscd}

\newlength{\minitwocolumn}
\makeatletter
\@addtoreset{equation}{section}
\makeatother

\newtheorem{thm}{Theorem}%[section]
\newtheorem{prop}[thm]{Proposition}

\newtheorem{cor}[thm]{Corollary}

\title{Bootstrap equations and 
correlation functions \\ for 
the Heisenberg XYZ antiferromagnet \\[2mm]
\large{(Short title: Bootstrap equations and 
XYZ correlation functions)}}

\author{Yas-Hiro Quano}

\date{\it Department of Clinical Engineering, 
Suzuka University of Medical Science \\
      \it Kishioka-cho, Suzuka 510-0293, Japan \\[5mm]
      \rm \normalsize{email: quanoy@suzuka-u.ac.jp}}
\begin{document}

\maketitle
\begin{abstract}
Presented are two kinds of integral solutions to the quantum 
Knizhnik-Zamolodchikov equations for the $2n$-point 
correlation functions of the Heisenberg XYZ 
antiferromagnet. Our first integral solution 
can be obtained from those for the cyclic SOS model 
by using the vertex-face correspondence. By the 
construction, the sum with respect to the local 
height variables $k_0 , k_1 , \cdots , k_{2n}$ 
of the cyclic SOS model remains other than $n$-fold 
integral in the first solution. In order to perform 
those summations, we improve that to find the second 
integral solution of $(r+1)n$-fold integral for 
$r\in \mathbb{Z}_{>1}$, 
where $r$ is a parameter of the XYZ model. 
Furthermore, we discuss the relations among our formula, 
Lashkevich-Pugai's formula and Shiraishi's one. \\[5mm] 
\end{abstract}

\section{Introduction}

Thirty years have passed since Baxter's papers on 
the eight-vertex model \cite{8V} and the XYZ spin chain 
\cite{XYZ} were published. In a series of papers 
\cite{3-bu/1,3-bu/2,3-bu/3} Baxter constructed the 
eigenvectors of the Hamiltonian of the XYZ spin chain 
by translating the eight-vertex model to the equivalent 
face model, the eight-vertex SOS model. 

Though many authors have tried to extract the results 
on correlation functions of the eight-vertex/XYZ model, 
very few results were obtained. The spontaneous polarization 
was conjectured by Baxter and Kelland \cite{BK}, and the 
same results were reproduced by solving a set of difference 
equations, the quantum Knizhnik-Zamolodchikov equation 
($q$-KZ equations) in \cite{JMN}. 
On the basis of the $Z$-invariance the integral formulae 
of correlation functions were obtained 
in the special case that the eight-vertex model can be 
factorized into two independent Ising model \cite{Zinv}. 
A free field representation of the type I \cite{LaP} 
and type II \cite{La} vertex operators were constructed 
to express the correlation functions and the form factors 
of the eight-vertex/XYZ model in terms of those of 
the eight-vertex SOS model \cite{LuP}. 

There are several ways to address the problem of 
correlation functions of integrable models. 
The vertex operator approach \cite{JMbk} provides 
a powerful tool to calculate the correlation functions 
of the six-vertex/XXZ model \cite{CORR}, the RSOS-type 
model \cite{LuP,AJMP,Ko}. The bootstrap approach 
\cite{JMN,FJMMN} is based on the corner transfer matrix (CTM) 
method \cite{ESM} and enables us to derive a set of difference 
equations of correlation functions. Note that the CTM method 
can be applied to massive integrable model such as the XYZ 
Heisenberg chain in the antiferromagnetic regime. 
Kitanin et al \cite{KMST1,KMST2} developed the 
algebraic Bethe Ansatz method to obtain domain wall 
correlation functions of the XXZ model, using the 
determinant structure of the partition function. 

In this paper we try to construct the correlation functions 
of the XYZ antiferromagnet by directly solving a set of 
difference equations, bootstrap equations. These equations 
are derived on the basis of CTM bootstrap in \cite{JMN}. 
In a previous paper \cite{bXXZ} we did the same thing 
for both the bulk and boundary XXZ antiferromagnet. 
Let us cite some results concerning 
the bulk XXZ case from \cite{bXXZ}. 

The $2n$-point correlation functions of the bulk 
XXZ antiferromagnet with the spectral parameters 
$(\zeta )=(\zeta_1 , \cdots , \zeta_{2n})$ and 
the spin variables $\varepsilon =(\varepsilon_1 , 
\cdots , \varepsilon_{2n})$ 
can be expressed up to a scalar function of 
$\zeta$'s in the following form: 
\begin{equation}
\overline{G}_{\sigma}^{(n)} 
(\zeta)^\varepsilon 
=\prod_{a\in A}\oint_{C_a} 
\dfrac{dw_a}{2\pi \sqrt{-1}w_a} 
\Psi_{\sigma}^{(n)} (\{w_a\}_{a\in A}| \zeta)
Q^{(n)}(\{w_a\}_{a\in A}| \zeta )^\varepsilon , 
~~~~ (\sigma =\pm ) 
\label{eq:G-form/XXZ}
\end{equation}
where 
\begin{equation}
A:=\{ a|\varepsilon_a >0, \,\,1\leqslant a\leqslant 2n \}. 
\label{eq:df-A/XXZ}
\end{equation}
The functions $Q^{(n)}(w| \zeta)^\varepsilon$ is a 
meromorphic function defined below, 
and $\Psi_{\sigma}^{(n)} (w| \zeta)$ is a kernel 
with appropriate transformation properties 
\cite[eq.(2.32--33)]{bXXZ} and recursion relations 
\cite[eq.(2.34)]{bXXZ}. The subscript $\sigma =\pm$ 
in (\ref{eq:G-form/XXZ}) specifies one of two vacuums 
of the XXZ model in the antiferromagnetic regime. 
The explicit expression of 
$Q^{(n)}(w| \zeta)^\varepsilon$ is as follows: 
\begin{equation}
Q^{(n)}(w|\zeta )^{\varepsilon} 
=\prod_{a\in A} \zeta_a \left( 
\prod_{j=1}^{a-1} (xz_j -w_a )
\prod_{j=a+1}^{2n} (xw_a -z_j ) \right)  
\prod_{a,b\in A\atop a<b} (x^{-1}w_b -xw_a )^{-1}, 
\label{eq:df-Q}
\end{equation}
where $z_j =\zeta_j^2$. Here we notice the relation 
between our previous work \cite{bXXZ} and 
\cite{massless-XXZ,JKM}, the latter in which 
the correlation functions of the massless XXZ model 
were obtained. Since the XXZ model has only one vacuum 
in the massless regime, the vacuum structure of 
the massless XXZ model is very different from that of 
the XXZ antiferromagnet. Nevertheless, the structures 
of the integral formulae for the correlation functions 
in the both regime are quite similar. In order to see 
such similarity, let us substitute $x=e^{-\epsilon}$, 
$\zeta_j =e^{-\epsilon\beta_j}$ and 
$w_a =e^{-2\epsilon\alpha_a}$ into (\ref{eq:df-Q}). 
Then we have 
\begin{equation}
\begin{array}{cl}
& Q^{(n)}(w|\zeta )^{\varepsilon} \\
=&{\rm const.} \displaystyle\prod_{a\in A} 
e^{(2n-1)\epsilon\alpha_a} \left( 
\prod_{j=1}^{a-1} {\rm sh}\,\epsilon (\alpha_a -
\beta_j -\tfrac{1}{2}) 
\prod_{j=a+1}^{2n} {\rm sh}\,\epsilon (\beta_j -
\alpha_a -\tfrac{1}{2}) \right) \\
\times&\displaystyle\displaystyle\prod_{j=1}^{2n} 
e^{n\epsilon\beta_j} \prod_{a,b\in A\atop a<b} \left( 
{\rm sh}\,\epsilon (\alpha_a -\alpha_b +1) \right)^{-1}. 
\end{array}
\label{eq:massless-Q}
\end{equation}
When $|x|=1$ ($\epsilon \in\sqrt{-1}\mathbb{R}$), 
the expression (\ref{eq:massless-Q}) is equal to 
the meromorphic function $Q_n (\alpha |\beta )^{\varepsilon}$ 
in \cite{massless-XXZ}, the massless analogue of 
$Q^{(n)}(w|\zeta )^{\varepsilon}$, 
up to a trivial factor. Furthermore, we should notice here 
that the corresponding expression for the eight-vertex 
SOS model which will be given in section 2 possesses also 
quite similar structure to (\ref{eq:massless-Q}). 

The XYZ Heisenberg antiferromagnet has two 
parameters $q$ and $p$, where $p=(-q)^r$. 
Lashkevich and Pugai \cite{LaP} wrote down the 
$n$-fold integral formulae for the $2n$-point 
correlation functions $G$ 
in terms of those of SOS model $F$ as follows: 
\begin{equation}
G(u_1 , \cdots , u_{2n}|u_0 )=
\sum_{k_0 , k_1 , \cdots k_{2n}} t^{k_0}_{k_1} (u_1 -u_0 ) 
\otimes \cdots \otimes t^{k_{2n-1}}_{k_{2n}} (u_{2n}-u_0 ) 
F(u_1 , \cdots , u_{2n}|u_0 )^{k_0 k_1 \cdots k_{2n}}. 
\label{eq:LaP-f}
\end{equation}
Here $u_1, \cdots , u_{2n}$ are additive spectral parameters, 
$u_0$ is an auxiliary parameter, 
and $t^k_{k\pm 1}(u)$ is the intertwining vectors 
introduced by Baxter \cite{3-bu/1,3-bu/2}. The sum with 
respect to $k_j$ ($1\leqslant j\leqslant 2n$) should be 
taken over $k_j =k_{j-1}\pm 1$. Concerning the sum with 
respect to $k_0$, that depends on the value of $r$. 
In (\ref{eq:LaP-f}) $r>1$ is supposed. When $r$ is a 
rational number of the form $2r=N/N'$ ($N$, $N'$ coprime), 
we should take the sum over $\mathbb{Z}/N\mathbb{Z}$, 
otherwise over $\mathbb{Z}$. 

On the other hand, Shiraishi \cite{Shi} constructed 
the formulae of the correlation functions of the XYZ 
model without using the vertex-face correspondence. 
There are three main differences between 
Lashkevich-Pugai's formulae and Shiraishi's ones. 
Shiraishi restricted himself to the case $r=\frac{3}{2}$, 
while $r>1$ in (\ref{eq:LaP-f}). Furthermore, the 
$2n$-point correlation functions \`{a} la Shiraishi 
are of the $2n$-fold integral form, while 
$n$-fold in the former case. These two differences 
seem disadvantage of Shiraishi's formulae in a sense, however, 
the third difference gives advantage. Shiraishi constructed 
the type II and type I vertex operators of the eight-vertex 
model as intertwiners of the $q$-deformed Virasoro algebra 
\cite{SKAO}, without transforming those of SOS model via 
vertex-face correspondence. Thus, the sum with respect to 
$k_j$ in (\ref{eq:LaP-f}) are not needed. 

In the present paper we wish to synthesize the advantages 
of the two formulae. Namely, we construct $(r+1)n$-fold 
integral formulae for the $2n$-point correlation of the XYZ 
model with $r\in\mathbb{Z}_{>1}$, 
by using the vertex-face correspondence, 
and performing the sum with respect to $k_j$. 

The rest of the present paper is organized as follows. 
In section 2 we formulate the XYZ antiferromagnet 
and the corresponding cyclic SOS model. 
In section 3 we present the bootstrap equations and 
construct integral formulae for correlation functions 
for the cyclic SOS model. In section 4 we further 
present integral formulae for correlation functions of 
the XYZ antiferromagnet. 
In section 5 we give some concluding remarks. In Appendix 
A we list some properties of the $R$-matrix of the 
eight-vertex model. 

\section{The XYZ antiferromagnet and the cyclic SOS model}

\subsection{The Hamiltonian and the $R$-matrix} 

The Jacobi theta functions 
with the characteristics $a,b\in \mathbb{R}$ are defined by 
\begin{equation}
\vartheta\left[\begin{array}{c} a \\ b \end{array} \right]
(u;\tau ): =\displaystyle\sum_{m\in \mathbb{Z}} 
\exp \left\{ \pi \sqrt{-1}(m+a)~\left[ (m+a)\tau 
+2(u+b) \right] \right\}. \label{Rieth}
\end{equation}
The $i$th theta functions are defined as follows: 
$$
\begin{array}{l}
\displaystyle\theta_1 (u;\tau )=
\vartheta\left[\begin{array}{c} 1/2 \\ -1/2 \end{array} \right]
(u;\tau ), ~~~~
\theta_2 (u;\tau )=
\vartheta\left[\begin{array}{c} 1/2 \\ 0 \end{array} \right]
(u;\tau ), \\
\displaystyle\theta_3 (u;\tau )=
\vartheta\left[\begin{array}{c} 0 \\ 0 \end{array} \right]
(u;\tau ), ~~~~\theta_4 (u;\tau )=
\vartheta\left[\begin{array}{c} 0 \\ 1/2 \end{array} \right]
(u;\tau ). \end{array}
$$
These theta functions have the infinite product forms, 
e.g., 
$$
\theta_1 (u;\tau )=\sqrt{-1}q^{\frac{1}{4}}
e^{-\sqrt{-1}\pi u} 
\Theta_{q^2} (e^{2\sqrt{-1}\pi u}), ~~~~ 
\theta_2 (u;\tau )=q^{\frac{1}{4}}
e^{-\sqrt{-1}\pi u} 
\Theta_{q^2} (-e^{2\sqrt{-1}\pi u}), 
$$
where $q=\exp (\sqrt{-1}\pi\tau )$ and 
we used the standard notation 
$$
\Theta_{p}(z):=(z; p)_\infty 
(pz^{-1}; p)_\infty (p, p)_\infty , ~~~~
(a;p_1,\cdots,p_n)_\infty=
\displaystyle \prod_{k_i\geqslant 0}(1-ap_1^{k_1}\cdots p_n^{k_n}). 
$$

In this section we consider the XYZ spin chain 
in an infinite lattice \cite{3-bu/3,ESM}
\begin{equation}
H_{XYZ}=-\frac{1}{2}\sum_{j\in\mathbb{Z}} 
\left( \sigma_{j+1}^x \sigma_{j}^x + 
\Gamma \sigma_{j+1}^y 
\sigma_{j}^y +\Delta \sigma_{j+1}^z \sigma_{j}^z \right). 
\label{eq:XYZ-H}
\end{equation}
Here $\sigma_{j}^x$, $\sigma_{j}^y$ and $\sigma_{j}^z$ 
denote the standard Pauli matrices acting on $j$-th site, 
and we 
restrict ourselves to the antiferromagnetic regime 
$|\Gamma |<1$ and $\Delta <-1$, where 
$$
\begin{array}{rcl}
\Gamma &=&\dfrac{1-\gamma}{1+\gamma}, 
~~~~\gamma =-\dfrac{\theta_1^2 
(\frac{\sqrt{-1}\epsilon}{\pi};
\frac{2\sqrt{-1}\epsilon r}{\pi})}{\theta_4^2 
(\frac{\sqrt{-1}\epsilon}{\pi};
\frac{2\sqrt{-1}\epsilon r}{\pi})}>0, \\[6mm]
\Delta &=&-\displaystyle\frac{1}{1+\gamma}
\frac{\theta_4^2 
(0;\frac{2\sqrt{-1}\epsilon r}{\pi})(\theta_2 
\theta_3 )(\frac{\sqrt{-1}\epsilon}{\pi};
\frac{2\sqrt{-1}\epsilon r}{\pi})}
{(\theta_2 \theta_3 )
(0;\frac{2\sqrt{-1}\epsilon r}{\pi})\theta_4^2 
(\frac{\sqrt{-1}\epsilon}{\pi};
\frac{2\sqrt{-1}\epsilon r}{\pi})}. 
\end{array}
$$
The condition $|\Gamma |<1$ and $\Delta <-1$ 
is equivalent to the one such that $\epsilon >0$ 
and $r>1$. The antiferromagnetic XXZ Hamiltonian can 
be obtained by taking the limit $r\rightarrow \infty$. 
In this limit we have $\gamma\rightarrow 0$ and 
$\Delta\rightarrow -(x+x^{-1})/2$, where 
$0<x=e^{-\epsilon}<1$. As well as known, thus obtained XXZ 
Hamiltonian commutes with the quantum affine algebra 
$U_{-x} \bigl( \widehat{\mathfrak{sl}_2}\bigr)$ \cite{JMbk}. 
Let $V=\mathbb{C}v_+ +\mathbb{C}v_-$ be a vector 
representation 
of $U_{-x} \bigl( \widehat{\mathfrak{sl}_2}\bigr)$. Then 
the limiting XXZ Hamiltonian formally acts on 
$V^{\otimes \infty}=\cdots \otimes V\otimes V\otimes 
\cdots$. In \cite{JMbk} the space of states 
$V^{\otimes \infty}$ was identified with the tensor 
product of level $1$ highest and level $-1$ lowest 
representations of 
$U_{-x} \bigl( \widehat{\mathfrak{sl}_2}\bigr)$. 
It is very likely that the same structure 
survives in the nonlimiting generic case. In other words, 
the XYZ antiferromagnet is expected to have the symmetry 
described by the elliptic affine algebra 
${\cal A}_{q,p} \bigl( \widehat{\mathfrak{sl}_2}\bigr)$ 
\cite{FIJKMY1,FIJKMY2}, where $q=-x$ and $p=x^{2r}$. 

The XYZ Hamiltonian (\ref{eq:XYZ-H}) 
can be obtained from the transfer matrix for the 
eight-vertex model, by taking 
logarithmic derivative with respect to the spectral 
parameter $\zeta$. The $R$-matrix 
$R_{8V}(\zeta) \in \mbox{End}(\mathbb{C}^2 
\otimes \mathbb{C}^2)$ of the eight-vertex model 
is given as follows: 
\begin{equation}
\begin{array}{rcl}
    R_{8V}(\zeta)&=&\frac{1}{\kappa (\zeta )} 
    \overline{R}_{8V}(\zeta), \\
    \overline{R}_{8V}(\zeta)&=&w_4 (u) I\otimes I 
    +w_2 (u) \sigma^x \otimes \sigma^x 
    +w_3 (u) \sigma^y \otimes \sigma^y 
    +w_1 (u) \sigma^z \otimes \sigma^z , 
    %\dfrac{1}{2}\sum_{\alpha =0}^3 \dfrac{\theta_{4-\alpha} 
%(\frac{1}{2r}-\frac{u}{r}; 
%\frac{\pi \sqrt{-1}}{\epsilon r})}{\theta_{4-\alpha} 
%(\frac{1}{2r}; \frac{\pi \sqrt{-1}}{\epsilon r})} 
%    \sigma^\alpha \otimes \sigma^\alpha , 
\end{array}
\label{eq:R-bar} 
\end{equation}
where $z=\zeta^2=x^{2u}$ 
and %$(\sigma^0 , \sigma^1 , \sigma^2 , \sigma^3 )=
%(I, \sigma^x , \sigma^y , \sigma^z )$. 
$$
w_\alpha (u)=\dfrac{\theta_{\alpha} 
(\frac{1}{2r}+\frac{u}{r}; 
\frac{\pi \sqrt{-1}}{\epsilon r})}{\theta_{\alpha} 
(\frac{1}{2r}; \frac{\pi \sqrt{-1}}{\epsilon r})}. 
$$
In what follows 
we denote $R_{8V}(\zeta)$ and 
$\overline{R}_{8V}(\zeta)$ by $R(\zeta)$ and 
$\overline{R}(\zeta)$, respectively, when there is 
no fear of confusion. In the above 
expression we modify Sklyanin's parametrization \cite{Skl1} 
such that (\ref{eq:R-bar}) coincides with Baxter's one 
\cite{ESM} in the principal regime. 
The normalization factor 
\begin{equation}
\begin{array}{rcl}
\kappa (\zeta)&=&
\dfrac{[u+1]}{[1]}\bar{\kappa }(\zeta ), ~~~~ 
{[}u{]}
=\displaystyle x^{\frac{u^2}{r}-u}\Theta_{x^{2r}}(x^{2u}); \\
\bar{\kappa }(\zeta )&=& \displaystyle\zeta^{\frac{r-1}{r}}
\frac{(x^4 z ; x^4 , x^{2r})_{\infty}
      (x^{2} z^{-1} ; x^4 , x^{2r})_{\infty}
      (x^{2r} z ; x^4 , x^{2r})_{\infty}
      (x^{2r+2} z^{-1} ; x^4 , x^{2r})_{\infty}}
     {(x^{4} z^{-1} ; x^4 , x^{2r})_{\infty}
      (x^2 z ; x^4 , x^{2r})_{\infty}
      (x^{2r} z^{-1} ; x^4 , x^{2r})_{\infty}
      (x^{2r+2}z ; x^4 , x^{2r})_{\infty}}, 
      \end{array}
\label{eq:kappa}
\end{equation}
is chosen such that the partition function per site 
is unity. In other words, the factor $\kappa (\zeta )$ 
is the partition function per site of the unnormalized 
model defined by $\overline{R}(\zeta)$. 
For later convenience we also introduce 
the symbol $\{ u\}$ by 
$$
\{ u\} =\displaystyle x^{\frac{u^2}{r}-u}
\Theta_{x^{2r}}(-x^{2u}). 
$$
Note that 
$$
\theta_{1} (\tfrac{u}{r}; 
\tfrac{\pi \sqrt{-1}}{\epsilon r})
=\sqrt{\tfrac{\epsilon r}{\pi}}\exp \left( 
-\tfrac{\epsilon r}{4} \right) [u], 
~~~~\theta_{4} (\tfrac{u}{r}; 
\tfrac{\pi \sqrt{-1}}{\epsilon r})
=\sqrt{\tfrac{\epsilon r}{\pi}}\exp \left( 
-\tfrac{\epsilon r}{4} \right) \{u\}. 
$$

Let $\mathbb{C}^2 =\mathbb{C}v_+ \oplus \mathbb{C}v_-$ 
and introduce the matrix elements of the $R$-matrix 
as follows: 
\begin{equation}
R(\zeta)v_{\varepsilon_1}\otimes v_{\varepsilon_2}=
\sum_{\varepsilon'_1,\varepsilon'_2 =\pm} 
v_{\varepsilon'_1}\otimes v_{\varepsilon'_2}
R(\zeta)_{\varepsilon_1 \varepsilon_2}
    ^{\varepsilon'_1 \varepsilon'_2}=
    \dfrac{1}{\kappa (\zeta )}
    \sum_{\varepsilon'_1,\varepsilon'_2 =\pm} 
v_{\varepsilon'_1}\otimes v_{\varepsilon'_2} 
    \overline{R}(\zeta)_{\varepsilon_1 \varepsilon_2}
    ^{\varepsilon'_1 \varepsilon'_2}. 
\label{eq:R-comp}
\end{equation}
Then the nonzero entries are given by 
\begin{equation}
\begin{array}{l}
\displaystyle\overline{R}(\zeta)^{++}_{++}
=\overline{R}(\zeta)^{--}_{--}=a(\zeta)=
\frac{(\theta_2 \theta_3) 
(\frac{u}{2r};\frac{\pi\sqrt{-1}}{\epsilon r}) 
(\theta_1 \theta_4) 
(\frac{u+1}{2r};\frac{\pi\sqrt{-1}}{\epsilon r})}
{(\theta_2 \theta_3) (0;\frac{\pi\sqrt{-1}}{\epsilon r}) 
(\theta_1 \theta_4) 
(\frac{1}{2r};
\frac{\pi\sqrt{-1}}{\epsilon r})},
\\[6mm]
\displaystyle\overline{R}(\zeta)^{+-}_{+-}
=\overline{R}(\zeta)^{-+}_{-+}=b(\zeta) 
=-\frac{(\theta_1 \theta_4) 
(\frac{u}{2r};
\frac{\pi\sqrt{-1}}{\epsilon r}) 
(\theta_2 \theta_3) (\frac{u+1}{2r};
\frac{\pi\sqrt{-1}}{\epsilon r})}
{(\theta_2 \theta_3) (0;\frac{\pi\sqrt{-1}}{\epsilon r}) 
(\theta_1 \theta_4) 
(\frac{1}{2r};
\frac{\pi\sqrt{-1}}{\epsilon r})},
\\[6mm]
\displaystyle \overline{R}(\zeta)^{+-}_{-+}
=\overline{R}(\zeta)^{-+}_{+-}=c(\zeta)
=\frac{(\theta_2 \theta_3) 
(\frac{u}{2r};\frac{\pi\sqrt{-1}}{\epsilon r}) 
(\theta_2 \theta_3) 
(\frac{u+1}{2r};\frac{\pi\sqrt{-1}}{\epsilon r})}
{(\theta_2 \theta_3) (0;\frac{\pi\sqrt{-1}}{\epsilon r}) 
(\theta_2 \theta_3) 
(\frac{1}{2r};
\frac{\pi\sqrt{-1}}{\epsilon r})},
\\[6mm]
\displaystyle \overline{R}(\zeta)^{++}_{--}
=\overline{R}(\zeta)^{--}_{++}=d(\zeta)
=-\frac{(\theta_1 \theta_4) 
(\frac{u}{2r};
\frac{\pi\sqrt{-1}}{\epsilon r}) 
(\theta_1 \theta_4) (\frac{u+1}{2r};
\frac{\pi\sqrt{-1}}{\epsilon r})}
{(\theta_2 \theta_3) (0;\frac{\pi\sqrt{-1}}{\epsilon r}) 
(\theta_2 \theta_3) 
(\frac{1}{2r};\frac{\pi\sqrt{-1}}{\epsilon r})}. 
\end{array}
\label{eq:abcd}
\end{equation}
Assume that the parameters $u$, $\epsilon$ and $r$ lie 
in the so-called principal regime \cite{ESM}: 
$$
\epsilon >0, ~~ r>1, ~~ -1<u<0;  ~~~~ 
0<p<x<\zeta^{-1}<1, 
$$
where $x=e^{-\epsilon}$, $p=x^{2r}$ and 
$\zeta=x^{u}$. 
The main properties of the $R$-matrix 
are listed in Appendix A. 

\subsection{Vertex-face correspondence}

In order to construct the eigenvectors of the eight-vertex 
model, Baxter introduced the following 
intertwining vector \cite{3-bu/1,3-bu/2}: 
\begin{equation}
\begin{array}{rcl}
t_{k\pm 1}^k (u)&=&f(u) \bar{t}_{k\pm 1}^k (u), \\
\bar{t}_{k\pm 1}^k (u)&=&(\sqrt{-1})^k (\theta_1 \theta_4 )
\left( 
\tfrac{k\mp u}{2r};\tfrac{\pi \sqrt{-1}}{\epsilon r}
\right)v_+ +(\sqrt{-1})^{-k} 
(\theta_2 \theta_3 )\left(
\tfrac{k\mp u}{2r};\tfrac{\pi \sqrt{-1}}{\epsilon r}
\right)v_- . 
\end{array}
\label{eq:int-vec}
\end{equation}
The normalization factor $f(u)$ satisfies the relation
\begin{equation}
C^2 [u] f(u)f(u-1)=1, ~~~~ 
C=\frac{\epsilon r}{\pi} e^{-\frac{\epsilon r}{4}} 
\frac{(x^{2r},x^{2r})_\infty}{(-x^{2r},x^{2r})_\infty}. 
\label{eq:C-df}
\end{equation}
The explicit expression of $f(u)$ is as follows: 
\begin{equation}
f(u)=\frac{x^{-\frac{u^2}{2r}+\frac{r-1}{2r}u+\frac{1}{4}}}
{C \sqrt{(x^{2r}; x^{2r})_\infty}} 
\frac{(x^4 z; x^4, x^{2r})_\infty 
(x^{2r+2}z; x^4, x^{2r})_\infty}
{(x^2 z; x^4, x^{2r})_\infty (x^{2r}z; x^4, x^{2r})_\infty}
\label{eq:f-def}
\end{equation}
The SOS model is a face model \cite{3-bu/2} which is 
defined on the square lattice with a site variable 
$k_i \in \mathbb{Z}$ attached to each site $i$. 
We call $k_i $ local state or height 
and impose the condition 
that heights of adjoining sites differ by one. 
Local Boltzmann weight are assumed 
to be the function of the spectral parameter $u$ and 
to be denoted by 
$\displaystyle W\left( \left. \begin{array}{cc} 
c & d \\ b & a \end{array} \right| u \right)$, 
which is given for a state configuration 
$\displaystyle \begin{array}{c} 
\raisebox{1mm}{$c$} \\ \raisebox{-2mm}{$b$} \end{array} 
\fbox{\rule[-3mm]{0cm}{6mm}~~~~~} \begin{array}{c} 
\raisebox{1mm}{$d$} \\ \raisebox{-2mm}{$a$} \end{array}$ 
round a face. 
Here the four states $a, b, c$ and $d$ are ordered clockwise 
from the SE corner. 
The Boltzmann weights of this model are assumed 
to be the function of the spectral parameter $u$ and 
the nonzero Boltzmann weights are given as follows: 
\begin{equation}
\begin{array}{rcl}
\displaystyle W\left( \left. \begin{array}{cc} 
k\pm 2 & k\pm 1 \\ k\pm 1 & k \end{array} \right| u \right)
& = & \dfrac{1}{\bar{\kappa }(u)}, \\
~ & ~ & ~ \\
\displaystyle W\left( \left. \begin{array}{cc} 
k & k\pm 1 \\ k\pm 1 & k \end{array} \right| u \right)
& = & \displaystyle \dfrac{1}{\bar{\kappa }(u)}
\frac{[1]\{k\mp u\}}{[u+1]\{k\}}, \\
~ & ~ & ~ \\
\displaystyle W\left( \left. \begin{array}{cc} 
k & k\mp 1 \\ k\pm 1 & k \end{array} \right| u \right)
& = & \displaystyle -\dfrac{1}{\bar{\kappa }(u)}
\frac{[u]\{k\pm 1\}}{[u+1]\{k\}}. 
\end{array}
\label{eq:HY}
\end{equation}
The normalization factor $\bar{\kappa}(u)$ given by 
(\ref{eq:kappa}) is chosen 
such that the partition function per cite is unity. 
Then we have 
the so-called vertex-face correspondence \cite{3-bu/2}: 
\begin{equation}
R(u_1 -u_2 ) 
t^{c}_{b}(u_1 )\otimes t^{b}_{a}(u_2 )=
\sum_{d} W\left( \left. \begin{array}{cc} 
c & d \\ b & a \end{array} \right| u_1 -u_2 \right) 
t^{d}_{a}(u_1 )\otimes t^{c}_{d}(u_2 ). 
\label{eq:JMO}
\end{equation}

The RHS's of the last two equations in (\ref{eq:HY}) 
contain the factors like $\{k\}$, while the corresponding 
Boltzmann weights of the ABF model \cite{ABF} are 
expressed in terms of the factors like $[k]$. Suppose 
that $r$ is a positive integer greater than $1$. 
Then the ABF model is called 
a restricted SOS (RSOS) model, since $[k]=0$ for 
$k\in r\mathbb{Z}$ and therefore the local states 
can be restricted as $k=1, \cdots , r-1$. 
On the other hand, 
the present face model (\ref{eq:HY}) is a cyclic 
SOS model because $\{ k+r \}=\{k\}$. In \cite{LaP} 
the RSOS type weights were used so that the following 
regularization was required. In order to 
avoid the pole resulting from $[k]$ in the denominator, 
$k\in \mathbb{Z}+\delta$ should be assumed with 
some real $\delta$, and the limit $\delta\rightarrow 0$ 
should be taken after all calculation \cite{LaP}. 
In our case we need not such 
regularization because 
$\{ k\}\neq 0$ for $k\in\mathbb{Z}$. 

We also notice that the intertwining 
vector (\ref{eq:int-vec}) is closely connected with 
$2r$ dimensional cyclic representation of Sklyanin 
algebra \cite{Skl2,HY}. Actually, the $L$-operator 
defined via $R_{8V}LL=LLR_{8V}$ relation 
can be factorized into the intertwining 
vector (\ref{eq:int-vec}) and its dual vector. 

\section{Bootstrap equations in the cyclic SOS model}

\subsection{Integral formulae for the cyclic SOS case} 

As preliminary, let us consider the bootstrap 
equations and correlation functions 
for the cyclic SOS model. Let 
\begin{equation}
F_\sigma^{(n)}(\zeta_1 , \cdots , \zeta_{2n})^{
kk_1 \cdots k_{2n-1}k} ~~~~ (\sigma =\pm)
\end{equation}
for $|k_j -k_{j-1}|=1$ ($1\leqslant j\leqslant 2n$) 
and $k_0 =k_{2n}=k$, be the $2n$-point 
correlation function of the eight-vertex SOS model. 
Then these functions satisfy the following three 
CTM bootstrap equations \cite{FJMMN}: 
\begin{equation}
\begin{array}{cl}
&F_{\sigma}^{(n)} 
(\cdots,\zeta_{j+1},\zeta_j,\cdots)^{\cdots 
k_{j-1}k_jk_{j+1} \cdots} \\
=&\displaystyle\sum_{k'_j} 
W\left( \left. \begin{array}{cc} 
k_{j+1} & k_j \\ k'_j & k_{j-1} \end{array} \right| 
\zeta_j/\zeta_{j+1} \right)
F_{\sigma}^{(n)} 
(\cdots,\zeta_j,\zeta_{j+1},\cdots)^{\cdots 
k_{j-1}k'_jk_{j+1} \cdots}. 
\end{array}
\label{eq:W-symm-comp} 
\end{equation}
\begin{equation}
F_{\sigma}^{(n)} 
(\zeta_1, \cdots , \zeta_{2n-1}, x^{2}\zeta_{2n})^{
k\cdots k_{2n-1}k}
=\sigma \tfrac{\{k\}}{\{k_{2n-1}\}} F_{\sigma}^{(n)} 
(\zeta_{2n}, \zeta_1, \cdots , \zeta_{2n-1})^{
k_{2n-1}k\cdots k_{2n-1}}. 
\label{eq:Fcyc-comp}
\end{equation}
\begin{equation}
\begin{array}{rcl}
F_{\sigma}^{(n)} (\zeta_1, \cdots , \zeta_{2n-2}, 
\zeta_{2n-1}, x^{-1}\zeta_{2n-1})^{
k\cdots k_{2n-2} k_{2n-1} k} &=& 
\frac{\delta_{k_{2n-2},k}}{\{k\}} 
F_{\sigma}^{(n-1)} 
(\zeta_1, \cdots , \zeta_{2n-2})^{k\cdots k} 
\end{array}
\label{eq:rec-F-comp}
\end{equation}
\begin{equation}
\begin{array}{rcl}
F_{\sigma}^{(n)} (\zeta_1, \cdots , \zeta_{2n-2}, 
\zeta_{2n-1}, -x^{-1}\zeta_{2n-1})^{
k\cdots k_{2n-2} k\pm 1 k} &=& \delta_{k_{2n-2},k}
\frac{e^{-\frac{\pi\sqrt{-1}}{2r}(1\pm 2k)}}
{\sqrt{-1}\{k\}} F_{\sigma}^{(n-1)} 
(\zeta_1, \cdots , \zeta_{2n-2})^{k\cdots k} 
\end{array}
\label{eq:rec-Fm-comp}
\end{equation}

Set 
\begin{equation}
 F_{\sigma}^{(n)}(\zeta)^{kk_1\cdots k_{2n-1}k}
=c_n \prod_{1\leqslant j< k \leqslant 2n} 
\zeta_j^{\frac{r-1}{r}} g(z_j/z_k)
\times \overline{F}_{\sigma}^{(n)}(\zeta)^{kk_1\cdots k_{2n-1}k}. 
\label{eq:G-bar}
\end{equation}
Here $c_n$ is a constant which will be determined below, 
and the function $g(z)$ has the properties 
\begin{equation}
g(z)=g(x^{-4}z^{-1}), ~~~~ 
\bar{\kappa} (\zeta )=\zeta^{\frac{r-1}{r}} \frac{g(z)}{g(z^{-1})}. 
\label{eq:g-prop}
\end{equation}
The explicit form of $g(z)$ is as follows: 
\begin{equation}
g(z)=\frac{(x^6 z;x^4,x^4,x^{2r})_{\infty}
(x^2 z^{-1};x^4,x^4,x^{2r})_{\infty}
(x^{2r+6} z;x^4,x^4,x^{2r})_{\infty}
(x^{2r+2} z^{-1};x^4,x^4,x^{2r})_{\infty}}
{(x^8 z;x^4,x^4,x^{2r})_{\infty}
(x^4 z^{-1};x^4,x^4,x^{2r})_{\infty}
(x^{2r+4} z;x^4,x^4,x^{2r})_{\infty}
(x^{2r} z^{-1};x^4,x^4,x^{2r})_{\infty}}. 
\label{eq:df-g}
\end{equation}

In order to present our integral formulae for 
$\overline{F}_{\sigma}^{(n)}(\zeta )$ let us 
prepare some notation. Let 
\begin{equation}
A:=\{ a|k_a =k_{a-1}+1, \,\,1\leqslant a\leqslant 2n \}. 
\label{eq:df-A}
\end{equation}
Then the number of elements of $A$ is equal to $n$, 
because we now set $k_{2n}=k_0 =k$. 
We often use the abbreviations 
$(w)=(w_{a_1}, \cdots , w_{a_n})$, 
$(w')=(w_{a_1}, \cdots , w_{a_{n-1}})$ and 
$(w'')=(w_{a_1}, \cdots , w_{a_{n-2}})$ for $a_j \in A$ 
such that $a_1 <\cdots <a_n$. Let us define the 
following meromorphic function 
\begin{equation}
\begin{array}{rcl}
Q^{(n)}(w|\zeta )^{kk_1\cdots k_{2n-1}k}
&=&\displaystyle\prod_{a\in A} 
\{v_a -u_a +\tfrac{1}{2}-k_a \} \left( 
\prod_{j=1}^{a-1} [v_a -u_j -\tfrac{1}{2}] 
\prod_{j=a+1}^{2n} [u_j -v_a - \tfrac{1}{2}] \right) \\
&\times& \displaystyle\prod_{j=1}^{2n-1} \{k_j \}^{-1}
\prod_{a,b\in A\atop a<b} 
[v_a -v_b +1]^{-1}, 
\end{array}
\label{eq:df-QF}
\end{equation}
where $w_a =x^{2v_a}$. Here we should notice that 
the structure of the expression (\ref{eq:df-QF}) is 
quite similar to (\ref{eq:massless-Q}). 

We wish to find integral formulae of the form 
\begin{equation}
\overline{F}_{\sigma}^{(n)} 
(\zeta)^{kk_1\cdots k_{2n-1}k}=\prod_{a\in A}\oint_{C_a} 
\dfrac{dw_a}{2\pi\sqrt{-1}w_a} 
\Psi_{\sigma}^{(n)} (w|\zeta )
Q^{(n)}(w|\zeta)^{kk_1\cdots k_{2n-1}k}. 
\label{eq:G-form}
\end{equation}
Here, the kernel has the form 
\begin{equation}
\Psi^{(n)}_\sigma (w| \zeta )=
\vartheta^{(n)}_\sigma (w | \zeta )
\prod_{a\in A}\prod_{j=1}^{2n} x^{-\frac{(v_a -u_j)^2}{2r}} 
\psi \Bigl(\frac{w_{a}}{z_j}\Bigr) 
\prod_{1\leqslant j<k\leqslant 2n} 
x^{-\frac{(u_j -u_k)^2}{4r}},
\label{eq:df-Psi}
\end{equation}
where
\begin{equation}
\psi(z)=
\frac{(x^{2r+3}z;x^4,x^{2r})_{\infty}
(x^{2r+3}z^{-1};x^4,x^{2r})_{\infty}}
{(xz;x^4,x^{2r})_{\infty}(xz^{-1};x^4,x^{2r})_{\infty}}. 
\label{eq:df-psi}
\end{equation}

For the function 
$\vartheta^{(n)}_\sigma(w|\zeta )$ 
we assume that
\begin{itemize}
\item
it is anti-symmetric and holomorphic
in the $w_a \in \mathbb{C} \backslash \{0\}$,
 
\item
it is symmetric and meromorphic
in the $\zeta_j\in \mathbb{C} \backslash \{0\}$,
 
\item
it has the two transformation properties 
\begin{eqnarray}
\displaystyle\frac
{\vartheta^{(n)}_\sigma (w |\zeta', x^2 \zeta_{2n})}
{\vartheta^{(n)}_\sigma (w | \zeta )}&=& 
\sigma x^{-2n+\frac{2n-1}{r}} \displaystyle\prod_{a\in A}w_{a}
\prod_{j=1}^{2n} \zeta_j^{-1}, 
\label{eq:z-sym}\\
\displaystyle \frac
{\vartheta^{(n)}_\sigma (w', x^4 w_{a_n}| \zeta )}
{\vartheta^{(n)}_\sigma (w | \zeta )}
&=& x^{-4n} \displaystyle\prod_{j=1}^{2n} 
\frac{z_j }{ w_{a_n}}.
\label{eq:x-sym}
\end{eqnarray}
\item it satisfies the following recursion relation
\begin{equation}
\begin{array}{cl}
&\displaystyle \frac{\vartheta_{\sigma}^{(n)}
(w', x^{-1}z_{2n-1} |\zeta'',\zeta_{2n-1}, 
\pm x^{-1} \zeta_{2n-1})}
{\vartheta_{\epsilon\sigma}^{(n-1)}(w' |\zeta'')} \\
=&\displaystyle (\pm x^{-1}\zeta_{2n-1}^2 )^{
-(n-\frac{1}{2})(1-\frac{1}{r})}
\prod_{j=1}^{2n-2} 
\zeta_j^{-\frac{r-1}{r}} 
\prod_{a\in A\atop a\neq a_{n}} (\pm 
x^{u_{2n-1}-v_a -\frac{1}{2}}) 
\Theta_{x^2}(x w_{a}/z_{2n-1}). 
\end{array}
\label{eq:th-rec}
\end{equation}
Here we also fix the constant $c_n$ as follows: 
\begin{equation}
c_n =\frac{(-1)^{n(n+1)/2}x^{n^2/2}}{
(x^2 ; x^2 )_\infty^{n(n-3)/2}
(x^{2r}; x^{2r})_\infty^{3n(n-1)/2}}\frac{
\{ x^2 \}_\infty^n \{ x^6 \}_\infty^n
\{ x^{2r+2} \}_\infty^n \{ x^{2r+6} \}_\infty^n}{
\{ x^4 \}_\infty^n \{ x^8 \}_\infty^n
\{ x^{2r} \}_\infty^n \{ x^{2r+4} \}_\infty^n}, 
\label{eq:c_n}
\end{equation}
\end{itemize}
where 
$$
\{ z \}_\infty :=(z ; x^4 , x^4 , x^{2r})_\infty . 
$$
The function $\vartheta_{\sigma}^{(n)}(w|\zeta )$
is otherwise arbitrary, and the choice of 
$\vartheta_{\sigma}^{(n)}(w|\zeta )$ 
corresponds to that of solutions to 
(\ref{eq:W-symm-comp}--\ref{eq:rec-Fm-comp}). 
The transformation property of 
$\vartheta_{\sigma}^{(n)}(w|\zeta )$ implies
\begin{eqnarray}
\frac{\Psi_{\sigma}^{(n)}(w|\zeta', x^2 \zeta_{2n})}
{\Psi_{\sigma}^{(n)}(w|\zeta )}
&=&\sigma\prod_{j=1}^{2n} 
\left( \frac{\zeta_{2n}}{\zeta_j}\right)^{\frac{r-1}{r}}
\prod_{a\in A}\frac{[v_a -u_{2n} -\frac{1}{2}]}
{[u_{2n}-v_a +\frac{3}{2}]}, 
\label{eq:trPsi1} \\
\frac{\Psi_{\sigma}^{(n)}(w',x^4 w_{a_n}|\zeta )}
{\Psi_{\sigma}^{(n)}(w|\zeta )}
&=&\prod_{j=1}^{2n}\frac{[u_{j}-v_{a_n}-\frac{1}{2}]}
{[v_{a_n}-u_{j}+\frac{3}{2}]}.
\label{eq:trPsi2}
\end{eqnarray}

The integrand may have poles at 
\begin{equation}
w_a =\left\{ 
\begin{array}{ll}
x^{\pm (1+4k+2rl)}z_j & (1\leqslant j\leqslant 2n, 
k,l\in \mathbb{Z}_{\geqslant 0}), \\
x^2 w_b & (b<a), \\
x^{-2} w_b & (b>a). 
\end{array} \right. 
\label{eq:pole-position}
\end{equation}
We choose the integration contour $C_a$ with 
respect to $w_a$ ($a\in A$) such that $C_a$ is along 
a simple closed curve oriented anti-clockwise, and 
encircles the points $x^{1+4k+2rl}z_j 
(1\leqslant j \leqslant 2n , 
k,l\in \mathbb{Z}_{\geqslant 0})$ 
and $x^2 w_b$ ($b<a$) but not $x^{-1-4k-2rl} z_j 
(1\leqslant j \leqslant 2n, 
k,l\in \mathbb{Z}_{\geqslant 0})$ 
nor $x^{-2} w_b$ ($b>a$). Thus the contour $C_a$ actually 
depends on $z_j$'s besides $a$, so that it should 
be denoted by $C_a (z)=C_a (z_1, \cdots , z_{2n})$, precisely. 
The LHS 
of (\ref{eq:Fcyc-comp}) refers to the analytic continuation with 
respect to $\zeta_{2n}$. 
Nevertheless, once we restrict ourselves 
to the principal regime $0<p<x<\zeta_j^{-1}<1$, 
we can tune all $C_a$'s to be the common integration 
contour $C: |w_a|=x^{-1}$ because of the inequality 
$xz_j <x^{-1}<x^{-1}z_j$. 

We are now in a position to state the following proposition 
regarding the correlation functions and the bootstrap 
equations in the cyclic SOS model case: 

\begin{prop} Assume the properties of 
the function $\vartheta^{(n)}_\sigma$ 
below (\ref{eq:df-psi}) and the integration contour 
$C_a$ below (\ref{eq:pole-position}). 
Then the integral formulae 
(\ref{eq:G-bar}, \ref{eq:G-form}) 
with (\ref{eq:df-g}, \ref{eq:df-QF}, 
\ref{eq:df-Psi}, \ref{eq:df-psi}) 
solves the three equations 
(\ref{eq:W-symm-comp}--\ref{eq:rec-Fm-comp}). 
\label{prop:qKZ}
\end{prop}

The proof of Proposition \ref{prop:qKZ} will be 
given in the subsequent subsections. 

\subsection{Proof of the $W$-symmetry} 

Let us first prove (\ref{eq:W-symm-comp}). For 
that purpose we have to consider four cases 
according as $k_j -k_{j-1}=\pm 1$ and 
$k_{j+1}-k_j =\pm 1$. 

Suppose that $k_j -k_{j-1}=k_{j+1}-k_j =-1$. Then the relation 
(\ref{eq:W-symm-comp}) holds, because the integrand 
$\overline{F}^{(n)}_\sigma (\zeta )$ is evidently 
symmetric with respect to $\zeta_j$ and $\zeta_{j+1}$. 

Let $(k_{j-1}, k_j , k_{j+1})=(m,m-1,m)$ for some $m$. 
Then the relation 
(\ref{eq:W-symm-comp}) reduces to 
\begin{equation}
\begin{array}{rcl}
\overline{F}_{\sigma}^{(n)} 
(\cdots,\zeta_{j+1},\zeta_j,\cdots)^{\cdots 
mm-1m \cdots} 
&=&\dfrac{[1]\{m+u_j -u_{j+1}\}}{[u_j -u_{j+1}+1]\{m\}}
\overline{F}_{\sigma}^{(n)} 
(\cdots,\zeta_j,\zeta_{j+1},\cdots)^{\cdots 
mm-1m \cdots} \\[2mm]
&-&\dfrac{[u_j -u_{j+1}]\{m+1\}}{[u_j -u_{j+1}+1]\{m\}}
\overline{F}_{\sigma}^{(n)} 
(\cdots,\zeta_j,\zeta_{j+1},\cdots)^{\cdots 
mm+1m \cdots}. 
\end{array}
\label{eq:W-symm/-+}
\end{equation}
Note that the set of the integration variables in 
the second term of the RHS is different from the 
other terms. Since $w_a =x^{2v_a}$'s are integration variables, 
we can replace both $v_{j}$ and $v_{j+1}$ in the integrand 
by $v$. After that, the relation (\ref{eq:W-symm/-+}) 
follows from the equality of the integrands. In this step 
we use 
$$
\begin{array}{rcl}
\{v-u_{j}+\frac{1}{2}-m\}[v-u_{j+1}-\frac{1}{2}]&=&
\dfrac{[1]\{m+u_j -u_{j+1}\}}{[u_j -u_{j+1}+1]\{m\}}
\{v-u_{j+1}+\frac{1}{2}-m\}[v-u_j -\frac{1}{2}] \\
&-&\dfrac{[u_j -u_{j+1}]\{m-1\}}{[u_j -u_{j+1}+1]\{m\}}
\{v-u_j -\frac{1}{2}-m\}[u_{j+1}-v-\frac{1}{2}]. 
\end{array}
$$

Suppose that $(k_{j-1},k_j,k_{j+1})=(m,m+1,m)$ for some 
$m$. This case can be 
proved in a similar way to the previous case. Here we use 
$$
\begin{array}{rcl}
\{v-u_{j+1}-\frac{1}{2}-m\}[u_{j}-v-\frac{1}{2}]&=&
\dfrac{[1]\{m-u_j +u_{j+1}\}}{[u_j -u_{j+1}+1]\{m\}}
\{v-u_j-\frac{1}{2}-m\}[u_{j+1}-v-\frac{1}{2}] \\
&-&\dfrac{[u_j -u_{j+1}]\{m+1\}}{[u_j -u_{j+1}+1]\{m\}}
\{v-u_{j+1}+\frac{1}{2}-m\}[v-u_j -\frac{1}{2}]. 
\end{array}
$$

Finally, let $(k_{j-1},k_j,k_{j+1})=(m-1,m,m+1)$ 
for some $m$. Then the integrand 
of the RHS of (\ref{eq:W-symm-comp}) contains the factor 
$$
I(u_j,u_{j+1};v_j,v_{j+1})=
\dfrac{\{v_j-u_{j}+\frac{1}{2}-m\}[u_{j+1}-v_j-\frac{1}{2}]
\{v_{j+1}-u_{j+1}-\frac{1}{2}-m\}[v_{j+1}-u_{j}-\frac{1}{2}]}
{[v_j-v_{j+1}+1]}. 
$$
The corresponding factor in the LHS should be equal to 
$I(u_{j+1},u_j ;v_j,v_{j+1})$. 
Thus the difference between both sides 
contains the factor 
$$
I(u_j,u_{j+1};v_j,v_{j+1})-
I(u_{j+1},u_{j};v_j,v_{j+1})
=\{m\}[u_{j}-u_{j+1}]\{v_j +v_{j+1}-u_j -u_{j+1}-m\}, 
$$
which is symmetric with respect to $w_j =x^{2v_j}$ and 
$w_{j+1}=x^{2v_{j+1}}$. Since 
$\Psi^{(n)}_\sigma (w|\zeta )$ is antisymmetric 
with respect to $w_a$'s, the relation 
(\ref{eq:W-symm-comp}) in this case does holds. 

\subsection{Proof of the cyclicity} 

In the proof of the cyclicity (\ref{eq:Fcyc-comp}) 
we have to consider the two cases $k_{2n-1}=k\pm 1$. 
First let $k_{2n-1}=k+1$. 
When the integral (\ref{eq:G-form}) 
is analytically continued from 
$\zeta_{2n}=x^{u_{2n}}$ to 
$x^2 \zeta_{2n}=x^{u_{2n}+2}$, the points 
$x^{1+4k+2rl}z_{2n}$ and $x^{-1-4k-2rl}z_{2n}$ 
($k,l \in\mathbb{Z}_{\geqslant 0}$) 
move to the points 
$x^{5+4k+2rl}z_{2n}$ and $x^{3-4k-2rl}z_{2n}$, 
respectively. 
In the LHS of (\ref{eq:Fcyc-comp}) the point $w_a =x^3 z_{2n}$ 
($v_a =u_{2n}+\frac{3}{2}$) are therefore 
outside the integral contour 
$C'_a=C_a (z', x^4 z_{2n})$. 
Nevertheless, we can deform $C'_a$ to the original one 
$C_a =C_a (z)$ 
without crossing any poles. That is because 
the factor 
$$
\prod_{a\in A} [u_{2n}-v_a +\tfrac{3}{2}], 
$$
contained in 
$Q^{(n)}(w|\zeta' , x^2 \zeta_{2n})^{k\cdots k+1k}$ 
cancels the singularity at $w_a =x^3 z_{2n}$. 
Thus the integral contours for both sides of 
(\ref{eq:Fcyc-comp}) coincide. 

Furthermore, by using (\ref{eq:trPsi1}) we obtain 
$$
\Psi^{(n)}_\sigma (w|\zeta', x^2 \zeta_{2n})
\prod_{a\in A} [v_a -u_{2n} -\tfrac{3}{2}]=\sigma
\Psi^{(n)}_\sigma (w|\zeta)
\prod_{a\in A} [u_{2n}-v_a +\tfrac{1}{2}] 
\prod_{j=1}^{2n-1} \left(\frac{\zeta_j}{\zeta_{2n}} 
\right)^{\frac{1}{r}}, 
$$
which implies that the integrands of both sides of 
(\ref{eq:Fcyc-comp}) coincide and therefore the relation 
(\ref{eq:Fcyc-comp}) holds when $k_{2n-1}=k+1$. 

Next let $k_{2n-1}=k-1$. 
In this case we make the rescale 
of variable $w_{2n}\mapsto x^4 w_{2n}$ 
($v_{2n}\mapsto v_{2n}+2$) in the LHS of 
(\ref{eq:Fcyc-comp}). Then the integral 
contour with respect to $w_a$ ($a\in A\backslash \{ 2n\}$) 
will be $C'_a =C_a (z', x^4 z_{2n})$, 
and the other one will be 
$\widetilde{C}=C_{2n}(x^{-4}z', z_{2n})$. 
For $a\in A\backslash \{ 2n\}$, we can deform the contour 
$C'_a$ to the original $C_a$ without crossing any poles, 
as the same reason as the previous case. The integral contour 
$\widetilde{C}$ encircles $x^{-3+4k+2rl}z_j$ 
and $x^{1+4k+2rl}z_{2n}$, but not 
$x^{-5-4k-2rl}z_j$ nor $x^{-1-4k-2rl}z_{2n}$, 
where $1\leqslant j\leqslant 2n-1$, 
$k,l\in \mathbb{Z}_{\geqslant 0}$. Since 
$Q^{(n)}(w|\zeta' , x^2 \zeta_{2n})^{k\cdots k-1k}$ 
contains the factor 
\begin{equation}
J(w'x^4 w_{2n}|\zeta',x^2\zeta_{2n})=
\{v_{2n}-u_{2n}+\tfrac{1}{2}-k\}
\prod_{j=1}^{2n-1} [v_{2n}-u_j+\tfrac{3}{2}]
\prod_{a\in A\atop a\neq 2n} 
\frac{[u_{2n}-v_a +\tfrac{3}{2}]}
{[v_a -v_{2n}-1]}, 
\label{eq:Qtr}
\end{equation}
the pole at $w_{2n}=x^{-3}z_{j}$ ($1\leqslant j\leqslant 2n-1$) 
disappears. Thus we can deform the contour 
$\widetilde{C}$ to the original one $C_{2n}=C_{2n}(z)$ 
without crossing any poles. Thus the integral contours 
in both sides of (\ref{eq:Fcyc-comp}) coincide. 

Replace the integral variables such that 
$(w', w_{2n})\mapsto (w_{2n}, w')$ in the RHS 
of (\ref{eq:Fcyc-comp}), and compare the integrands of both 
sides. Note that (\ref{eq:Qtr}) describes all $w_{2n}$- and 
$z_{2n}$-dependence of 
$Q^{(n)}(w|\zeta', x^2 \zeta_{2n})^{k\cdots k-1k}$. 
From (\ref{eq:trPsi1}, \ref{eq:trPsi2}) and the 
antisymmetric property of $\Psi^{(n)}_\sigma$ 
with respect to $w_a$'s, we have 
\begin{equation}
\Psi^{(n)}_\sigma (w', x^4 w_{2n}|
\zeta', x^2 \zeta_{2n})J(w'x^4 w_{2n}|\zeta',x^2\zeta_{2n})
=\sigma \Psi^{(n)}_\sigma 
(w_{2n}, w'|\zeta ) J(w_{2n}, w'|\zeta )
\prod_{j=1}^{2n-1} \left( \frac{\zeta_j}{\zeta_{2n}} 
\right)^{\frac{1}{r}}, 
\label{eq:Psi12}
\end{equation}
which implies that the integrands of both sides of 
(\ref{eq:Fcyc-comp}) coincide and therefore the relation 
(\ref{eq:Fcyc-comp}) holds when $k_{2n-1}=k-1$. 

\subsection{Proof of the normalization condition} 

Let us prove (\ref{eq:rec-F-comp}) and (\ref{eq:rec-Fm-comp}). 
The factor $g(z_{2n-1}/z_{2n})$ has a zero at 
$\zeta_{2n}=\pm x^{-1}\zeta_{2n-1}$. 
On the other hand, 
the two points $xz_{2n}$ and $x^{-1}z_{2n-1}$ are 
required to locate opposite sides of the integral 
contour $C_a$, so that a pinching may occur as 
$\zeta_{2n}\rightarrow\pm x^{-1}\zeta_{2n-1}$. 
If there is no pinching the correlation function 
$F^{(n)}_\sigma (\zeta )^{k\cdots k}$ vanishes at 
$\zeta_{2n}=\pm x^{-1}\zeta_{2n-1}$. 

Suppose that $(k_{2n-2}, k_{2n-1}, 
k_{2n})=(k+2, k+1, k)$. In this case 
the factor $[v_a -u_{2n-1}+\frac{1}{2}]$ contained in 
$Q^{(n)}(w|\zeta)^{k\cdots k+2 k+1 k}$ cancel the poles of 
$\psi (w_a /z_{2n-1})$ at $w_a =x^{-1}z_{2n-1}$, 
and therefore no pinching occurs as 
$\zeta_{2n}\rightarrow\pm x^{-1}\zeta_{2n-1}$. 
Thus the conditions (\ref{eq:rec-F-comp}) and 
(\ref{eq:rec-Fm-comp}) hold 
when $(k_{2n-2}, k_{2n-1}, k_{2n})=(k+2, k+1, k)$. 

When $(k_{2n-2}, k_{2n-1}, k_{2n})=(k-2, k-1, k)$, 
no pinching occurs for the 
integral contour for $C_a$ ($a\in A\backslash 
\{ 2n-1, 2n\}$), as the same reason as the previous case. 
Concerning the integral with respect to $w_{2n-1}$ 
and $w_{2n}$, there is no singularities 
at $w_{2n-1}=xz_{2n}$ and $w_{2n}=xz_{2n}$ and 
consequently no pinching occurs actually. 
In order to see such vanishing singularities, 
we first notice that the function 
$Q^{(n)}(w|\zeta )^{k\cdots k-2 k-1 k}$ contains the factor: 
$$
\begin{array}{cl}
&q(v_{2n-1}, v_{2n}; u_{2n-1}, u_{2n})\\
=&\dfrac{\{v_{2n-1}-u_{2n-1}+\frac{3}{2}-k\}
[u_{2n}-v_{2n-1}-\frac{1}{2}]
\{v_{2n}-u_{2n}+\frac{1}{2}-k\}
[v_{2n}-u_{2n-1}-\frac{1}{2}]}
{[v_{2n-1}-v_{2n}+1]}. 
\end{array}
$$
Because of the antisymmetric property of 
$\vartheta^{(n)}_\sigma$ with respect to $w_a$'s, 
the zero of 
$\vartheta^{(n)}_\sigma (w'', w_{2n-1}, xz_{2n}|\zeta )$ 
at $w_{2n-1}=xz_{2n}$ cancels the poles of 
$\psi (w_{2n-1}/z_{2n})$ at the same point. 
Furthermore, the poles of $\psi (w_{2n}/z_{2n})$ 
and $\psi (w_{2n}/z_{2n-1})$ at $w_{2n}=xz_{2n}$ 
and $z_{2n}=x^{-2}z_{2n-1}$ are canceled by 
the zeros of $g(z_{2n-1}/z_{2n})$ and 
$Q^{(n)}(w'', w_{2n-1}, xz_{2n}|\zeta )^{k\cdots k-2 k-1 k}$ 
at the same points. The latter cancellation can be shown 
as follows. 
Note that the integral is invariant as we replace 
$Q^{(n)}(w'', w_{2n-1}, xz_{2n}|\zeta )^{
k\cdots k-2 k-1 k}$ 
by its antisymmetric part with respect to 
$w_{2n-1}$ and $xz_{2n}$. Accordingly, we can replace 
the factor $q(v_{2n-1}, u_{2n}+\frac{1}{2}; u_{2n-1}, u_{2n})$ 
by the following factor: 
$$
\begin{array}{cl}
&q(v_{2n-1}, u_{2n}+\tfrac{1}{2}; u_{2n-1}, u_{2n})
-q(u_{2n}+\tfrac{1}{2}, v_{2n-1}; u_{2n-1}, u_{2n})\\
=& 
\{v_{2n-1}-u_{2n-1}+\frac{3}{2}-k\}
\{1-k\}[u_{2n-1}-u_{2n}]+\frac{\{u_{2n}-u_{2n-1}+2-k\}[1]
\{v_{2n-1}-u_{2n}+\tfrac{1}{2}-k\}
[v_{2n-1}-u_{2n-1}-\tfrac{1}{2}]}
{[u_{2n}+\tfrac{3}{2}-v_{2n-1}]}, 
\end{array}
$$
that vanishes when $z_{2n}=x^{-2}z_{2n-1}$ 
($u_{2n}=u_{2n-1} -1$ or $u_{2n}=u_{2n-1} -1-
\frac{\pi\sqrt{-1}}{\epsilon}$). 
Thus there is no singularity at $w_{2n}=xz_{2n}$ as 
$\zeta_{2n}\rightarrow\pm x^{-1}\zeta_{2n-1}$. 
The same thing at $w_{2n-1}=xz_{2n}$ can be easily shown 
in the same way. Hence the conditions (\ref{eq:rec-F-comp}) 
and (\ref{eq:rec-Fm-comp}) are verified when 
$(k_{2n-2}, k_{2n-1}, k_{2n})=(k-2, k-1, k)$. 

Next let 
$(k_{2n-2}, k_{2n-1}, k_{2n})=(k, k-1, k)$ 
and consider the limit $\zeta_{2n}\rightarrow x^{-1}
\zeta_{2n-1}$. 
In this case there is no pinching for the integrals 
with respect to $w_a$ ($a\in A\backslash \{ 2n \}$). 
Let $\widehat{C}$ denote the integral contour with 
respect to $w_{2n}$ such that $\widehat{C}$ encircles 
the same points as $C_{2n}$ does but $xz_{2n}$. Note that 
no pinching occurs with respect to the integral along 
$\widehat{C}$ because both the two points $x^{-1}z_{2n-1}$ 
and $xz_{2n}$ lie outside the contour $\widehat{C}$. 
Thus the integral with respect to $w_{2n}$ along 
the contour $C_{2n}$ can be replaced by the 
residue at $w_{2n}=xz_{2n}$. 

In order to evaluate the residue, the following formulae 
are useful: 
\begin{equation}
\begin{array}{cl}
&\displaystyle\lim_{z_{2n}\rightarrow x^{-2}z_{2n-1}} 
g\left( \frac{z_{2n-1}}{z_{2n}} \right) 
\psi \left( \frac{xz_{2n}}{z_{2n-1}} \right) \\[4mm]
=& \dfrac{1}
{(x^2; x^{2r})_\infty (x^{2r-2}; x^{2r})_\infty}
\dfrac{\{ x^4 \}_\infty \{ x^8 \}_\infty
\{ x^{2r} \}_\infty \{ x^{2r+4} \}_\infty}
{\{ x^2 \}_\infty \{ x^6 \}_\infty
\{ x^{2r+2} \}_\infty \{ x^{2r+6} \}_\infty}, 
\end{array}
\label{eq:lim_z}
\end{equation}
\begin{equation}
{\rm Res}_{w_{2n}=xz_{2n}}\frac{dw_{2n}}{w_{2n}}
\psi \left( \frac{w_{2n}}{z_{2n}} \right) 
=\frac{1}{(x^2; x^2 )_\infty (x^{2r}; x^{2r})_\infty}. 
\label{eq:Res_w}
\end{equation}
\begin{equation}
g(z)g(x^2 z)\psi (xz)[u+1] =
x^{\frac{(u+1)^2}{r}-(u+1)}(x^{2r}; x^{2r})_\infty . 
\label{eq:ggpsi_j}
\end{equation}
\begin{equation}
\psi \left( \frac{w}{z} \right) 
\psi \left( \frac{x^2 w}{z} \right) 
[u-v-\tfrac{1}{2}] =\frac{x^{\frac{(u-v-1/2)^2}{r}-(u-v-1/2)}
(x^{2r}; x^{2r})_\infty}
{(x w/z; x^2)_\infty(xz/w; x^2)_\infty}. 
\label{eq:ppQ_a}
\end{equation}
Using these, the condition (\ref{eq:rec-F-comp}) 
with $(k_{2n-2}, k_{2n-1}, k_{2n})=(k,k-1,k)$ reduces to 
$$
\begin{array}{rcl}
c_{n-1} \vartheta_{\sigma}^{(n-1)}(w'|\zeta'')
&=&c_n \vartheta_\sigma^{(n)}(w', x^{-1}z_{2n-1}|
\zeta'', \zeta_{2n-1}, x^{-1}\zeta_{2n-1}) \\
&\times& \displaystyle (-1)^{n} 
x^{-(2n-1)(1-\frac{1}{2r})} 
\zeta_{2n-1}^{(2n-1)(1-\frac{1}{r})} 
(x^{2}; x^{2})_\infty^{n-2} 
(x^{2r}; x^{2r})_\infty^{3(n-1)} \\
&\times&\displaystyle
\frac{\{ x^4 \}_\infty \{ x^8 \}_\infty
\{ x^{2r} \}_\infty \{ x^{2r+4} \}_\infty}
{\{ x^2 \}_\infty \{ x^6 \}_\infty
\{ x^{2r+2} \}_\infty \{ x^{2r+6} \}_\infty} 
\prod_{j=1}^{2n-2} \zeta_j^{\frac{r-1}{r}} 
%\\ &\times&\displaystyle 
\prod_{a\in A\atop a\neq 2n} 
\frac{x^{v_a -u_{2n-1} +\frac{1}{2}}}
{\Theta_{x^2}(xw_a /z_{2n-1})}, 
\end{array}
$$
which is valid under the assumption of 
(\ref{eq:th-rec}) and ({\ref{eq:c_n}). 
The relation (\ref{eq:rec-Fm-comp}) with 
$(k_{2n-2}, k_{2n-1}, k_{2n})=(k,k+1,k)$ 
can be similarly proved. 

When $(k_{2n-2}, k_{2n-1}, k_{2n})=(k,k+1,k)$, 
the only difference 
from the previous case is that the rational function 
$Q^{(n)}(w|\zeta )^{\cdots k k+1 k}$ contains the factor 
\begin{equation}
\left. \frac{\{v_{2n-1}-u_{2n-1}-\tfrac{1}{2}-k\}
[u_{2n}-v_{2n-1}-\tfrac{1}{2}]}{\{k+1\}}
\right|_{w_{2n-1}=xz_{2n}}=-
\frac{\{u_{2n}-u_{2n-1}-k\}[1]}{\{k+1\}}, 
\label{eq:Q+-}
\end{equation}
in the present case, while the corresponding factor 
in the previous case is 
\begin{equation}
\left. \frac{\{v_{2n}-u_{2n}+\tfrac{1}{2}-k\}
[v_{2n}-u_{2n-1}-\tfrac{1}{2}]}{\{k-1\}}
\right|_{w_{2n}=xz_{2n}}=
[u_{2n}-u_{2n-1}]. 
\label{eq:Q-+}
\end{equation}
Since (\ref{eq:Q+-}) is equal to 
(\ref{eq:Q-+}) as 
$\zeta_{2n}=x^{-1}\zeta_{2n-1}$, the condition 
(\ref{eq:rec-F-comp}) with 
$(k_{2n-2}, k_{2n-1}, k_{2n})=(k,k+1,k)$ 
follows from the previous 
case. The condition (\ref{eq:rec-Fm-comp}) with 
$(k_{2n-2}, k_{2n-1}, k_{2n})=(k,k+1,k)$ 
can be similarly proved. 
After all, the conditions 
(\ref{eq:rec-F-comp}) and (\ref{eq:rec-Fm-comp}) 
were componentwisely proved. 

\subsection{Non-trivial theta function}

In subsections 3.2--3.4 we proved Proposition 
\ref{prop:qKZ}. 
The key point in proving (\ref{eq:W-symm-comp}) was 
the $W$-matrix symmetry of the rational function 
$Q^{(n)}(w|\zeta )^{k\cdots k}$. We observed that 
the other two conditions (\ref{eq:Fcyc-comp}) and 
(\ref{eq:rec-F-comp}) hold under the assumption of 
the transformation properties and 
the recursion relation of $\vartheta^{(n)}_\sigma$. 
Actually, you can easily find that (\ref{eq:z-sym}), 
(\ref{eq:x-sym}) and (\ref{eq:th-rec}) are sufficient 
conditions of (\ref{eq:Fcyc-comp}) for $k_{2n-1}=k+1$, 
(\ref{eq:Fcyc-comp}) for $k_{2n-1}=k-1$ and 
(\ref{eq:rec-F-comp}), respectively. 

In this way we obtained the integral solutions to 
CTM bootstrap equations for correlation 
functions of the cyclic SOS model, with the 
freedom of the choice of $\vartheta^{(n)}_\sigma$. 
Now we wish to present an example of 
$\vartheta^{(n)}_\sigma$ satisfying 
all the properties given below (\ref{eq:df-psi}). 
\begin{equation}
\vartheta^{(n)}_\sigma (w|\zeta )
=\Theta_{x^2} \left( -\sigma x\prod_{a\in A} w_a^{-1} 
\prod_{j=1}^{2n}\zeta_j \right) 
\prod_{j=1}^{2n}\zeta_j^{-(n-\frac{1}{2})(1-\frac{1}{r})} 
\prod_{a,b\in A\atop a<b} x^{v_b -v_a} 
\Theta_{x^2} (w_a /w_b ). 
\label{eq:th-sol}
\end{equation}
You can easily see that (\ref{eq:th-sol}) satisfies 
all the properties of symmetry with respect to 
$\zeta_j$'s, antisymmetry with respect to $w_a$'s, 
(\ref{eq:z-sym}), (\ref{eq:x-sym}) and (\ref{eq:th-rec}). 
We also notice that there is actually no poles at 
$w_a =x^2 w_b$ ($b<a$) and $w_a =x^{-2}w_b$ ($b>a$) 
when we fix $\vartheta^{(n)}_\sigma$ to (\ref{eq:th-sol}) 
because of the factor $\Theta_{x^2} (w_a /w_b )$. 
Furthermore, we should like to notice that 
this example of the function 
$\vartheta^{(n)}_\sigma (w|\zeta )$ is quite similar to 
the anti-ferromagnetic XXZ analogue \cite{bXXZ}. 

\section{Integral formulae for the XYZ antiferomagnet}

\subsection{Correlation functions and difference equations}

Let us introduce the $V^{\otimes 2n}$-valued 
correlation functions 
\begin{equation}
G^{(n)}_\sigma(\zeta_1,\cdots,\zeta_{2n})=
\sum_{\varepsilon_j =\pm} v_{\varepsilon_1} \otimes 
\cdots \otimes v_{\varepsilon_{2n}} 
G^{(n)}_\sigma(\zeta_1,\cdots,\zeta_{2n})
^{\varepsilon_1 \cdots \varepsilon_{2n}}, 
\label{eq:df-corr}
\end{equation}
where $\sigma =\pm$ signifies one of two vacuums of 
the XYZ Heisenberg antiferromagnet. Let 
\begin{equation}
{\cal O}=E^{(1)}_{\varepsilon_1 \varepsilon'_1} 
\cdots E^{(n)}_{\varepsilon_n \varepsilon'_n}, 
\label{eq:l-op}
\end{equation}
where $E^{(j)}_{\varepsilon_j \varepsilon'_j}$ 
is the matrix unit on the $j$-th site. 
Then the correlation 
function (\ref{eq:df-corr}) gives the expectation 
value of the local operator (\ref{eq:l-op}) 
by specializing the spectral parameters as follows: 
\begin{equation}
\langle {\cal O} \rangle_\sigma 
=G^{(n)}_\sigma(\overbrace{x^{-1}\zeta,\cdots,
x^{-1}\zeta}^{n}, \overbrace{\zeta,\cdots,\zeta}^{n})^{
-\varepsilon_{n}\cdots -\varepsilon_{1}\varepsilon'_1 
\cdots \varepsilon'_{n}}. 
\label{eq:spec}
\end{equation}

In what follows we often use the abbreviations: 
$(\zeta) =(\zeta_1 , \cdots , \zeta_{2n})$, 
$(\zeta') =(\zeta_1 , \cdots , \zeta_{2n-1})$, 
$(\zeta'') =(\zeta_1 , \cdots , \zeta_{2n-2})$, 
$(z)=(z_1 , \cdots z_{2n})$, 
$(z')=(z_1 , \cdots z_{2n-1})$; and 
$(\varepsilon )=(\varepsilon_1 \cdots \varepsilon_{2n})$, 
$(\varepsilon')=(\varepsilon_1 \cdots \varepsilon_{2n-1})$, 
$(\varepsilon'')=(\varepsilon_1 \cdots \varepsilon_{2n-2})$. 
On the basis of the CTM (corner transfer matrix) bootstrap 
approach, the correlation functions satisfy the following 
three conditions \cite{JMN}: 

\noindent{\it 1. $R$-matrix symmetry}
\begin{equation}
P_{j\,j+1} G_{\sigma}^{(n)} 
(\cdots,\zeta_{j+1},\zeta_j,\cdots) 
\quad =
R_{j\,j+1}(\zeta_j/\zeta_{j+1})G_{\sigma}^{(n)} 
(\cdots,\zeta_j,\zeta_{j+1},\cdots)
\qquad (1\leqslant j\leqslant 2n-1), 
\label{eq:R-symm} 
\end{equation}
where $P(x\otimes y)=y\otimes x$. 

\noindent{\it 2. Cyclicity}
\begin{equation}
P_{12}\cdots P_{2n-1 2n} 
G_{\sigma}^{(n)} (\zeta', x^{2}\zeta_{2n})
=\sigma G_{\sigma}^{(n)} (\zeta_{2n}, \zeta'). 
\label{eq:cyc}
\end{equation}

\noindent{\it 3. Normalization}
\begin{equation}
\begin{array}{rcl}
G_{\sigma}^{(n)} 
(\zeta'',\zeta_{2n-1}, \zeta_{2n})|_{
\zeta_{2n}=s x^{-1}\zeta_{2n-1}} &=&
G_{s\sigma}^{(n-1)} (\zeta'')
\otimes u_s  ~~~~ (s =\pm ), 
\end{array}
\label{eq:rec-G}
\end{equation}
where $u_s =v_+ \otimes v_- +s v_- \otimes v_+$. 

These three conditions can be componentwisely recast 
as follows: 
\begin{equation}
\begin{array}{cl}
&G_{\sigma}^{(n)} 
(\cdots,\zeta_{j+1},\zeta_j,\cdots)^{\cdots 
\varepsilon_{j+1}\varepsilon_j \cdots} \\
=&\displaystyle\sum_{
\varepsilon'_j, \varepsilon'_{j+1}=\pm} 
R(\zeta_j/\zeta_{j+1})^{\varepsilon_j 
\varepsilon_{j+1}}_{\varepsilon'_j \varepsilon'_{j+1}}
G_{\sigma}^{(n)} 
(\cdots,\zeta_j,\zeta_{j+1},\cdots)^{\cdots 
\varepsilon'_{j}\varepsilon'_{j+1} \cdots}. 
\end{array}
\label{eq:R-symm-comp} 
\end{equation}
\begin{equation}
G_{\sigma}^{(n)} 
(\zeta', x^{2}\zeta_{2n})^{\varepsilon' \varepsilon_{2n}}
=\sigma G_{\sigma}^{(n)} 
(\zeta_{2n}, \zeta')^{\varepsilon_{2n}\varepsilon'}. 
\label{eq:cyc-comp}
\end{equation}
\begin{equation}
\begin{array}{rcl}
G_{\sigma}^{(n)} 
(\zeta'',\zeta_{2n-1}, sx^{-1}\zeta_{2n-1})^{
\varepsilon''\, \varepsilon_{2n-1}\varepsilon_{2n}} 
&=&\varepsilon_{2n-1}^{(1-s )/2}
\delta_{\varepsilon_{2n-1}+\varepsilon_{2n},0}
G_{s\sigma}^{(n-1)} (\zeta'')^{\varepsilon''} 
\;\;\;\; (s=\pm ). 
\end{array}
\label{eq:rec-G-comp}
\end{equation}
Combining (\ref{eq:R-symm-comp}) and (\ref{eq:rec-G-comp}) 
we obtain another expression of the normalization 
condition: 
$$
\displaystyle\sum_{s=\pm}
\varepsilon^{(1-s)/2} G_{\sigma}^{(n)} 
(\zeta'',\zeta_{2n-1}, s x\zeta_{2n-1})^{
\varepsilon''\, \varepsilon\, -\varepsilon}=
G_{s\sigma}^{(n-1)} (\zeta'')^{\varepsilon''}, 
$$
where we also use 
$$
R(s x^{-1})=\left( \begin{array}{cccc} 
0 & 0 & 0 & 0 \\
0 & s & 1 & 0 \\ 0 & 1 & s & 0 \\ 
0 & 0 & 0 & 0 \end{array} \right). 
$$

Note that the first 
two conditions imply the difference equation of 
the quantum KZ type \cite{FR} of level $-4$: 
\begin{equation}
\begin{array}{l}
T_j G_{\sigma}^{(n)} (\zeta)
=
R_{j\,j-1}(x^{-2} \zeta_j/\zeta_{j-1})
\cdots
R_{j\,1}(x^{-2} \zeta_{j}/\zeta_1 ) \\
\qquad\qquad \times
R_{j\,2n}(\zeta_{j}/\zeta_{2n})
\cdots
R_{j\,j+1}(\zeta_{j}/\zeta_{j+1})
G_{\sigma}^{(n)} (\zeta), 
\label{eq:qKZ}
\end{array}
\end{equation}
where $T_j$ is the shift operator such that 
\begin{equation}
T_j F (\zeta)
=F (\zeta_1,\cdots,x^{-2}\zeta_j,\cdots,\zeta_{2n}), 
\label{eq:T-shift}
\end{equation}
for any $2n$-variate function $F$. When the definition 
of $T_j$ is replaced by 
$$
T_j F (\zeta)
=F (\zeta_1,\cdots,x^{l+2}\zeta_j,\cdots,\zeta_{2n}), 
$$
and all the arguments $x^{-2}\zeta_j /\zeta_k$ in the 
first line of the RHS of (\ref{eq:qKZ}) are also replaced 
by $x^{l+2}\zeta_j /\zeta_k$, the difference equation 
(\ref{eq:qKZ}) is called the quantum KZ equations of 
level $l$. 

{\bf Remark.} As a result of $\mathbb{Z}_2$-symmetry 
of $R$-matrix, there are two ground states in 
the XYZ antiferromagnet. Let us specify the 
two ground states by $i=0, 1$, and denote the correlation 
function on the $i$-th ground state by $G^{(n)}_i (\zeta)$. 
Representation theoretical speaking, the correlation 
function $G^{(n)}_i (\zeta)$ refers to the trace of type I 
vertex operators on the irreducible highest weight module 
${\cal H}^{(i)}$ of 
${\cal A}_{q,p} \bigl( \widehat{\mathfrak{sl}_2}\bigr)$ 
\cite{FIJKMY2}. 
The CTM bootstrap approach suggest that both 
$G^{(n)}_0 (\zeta)$ and $G^{(n)}_1 (\zeta)$ will appear 
in the cyclicity condition as follows: 
$$
P_{12}\cdots P_{2n-1 2n} G_{i}^{(n)} 
(\zeta_1,\cdots,\zeta_{2n-1}, x^{2}\zeta_{2n})
=G_{1-i}^{(n)} 
(\zeta_{2n}, \zeta_1,\cdots,\zeta_{2n-1}). 
$$
Thus we introduce $G^{(n)}_\sigma(\zeta)
=G^{(n)}_0 (\zeta)+\sigma G^{(n)}_1 (\zeta)$ 
such that the second equation (\ref{eq:cyc}) 
involves only $G^{(n)}_\sigma (\zeta)$. 

Thanks to (\ref{eq:g-prop}) the first two equations 
(\ref{eq:R-symm-comp}--\ref{eq:cyc-comp}) are 
rephrased in terms of 
$\overline{G}_{\sigma}^{(n)}(\zeta)$ 
and $\overline{R}(\zeta)$ as follows 
\begin{equation}
\overline{G}_{\sigma}^{(n)} 
(\cdots,\zeta_{j+1},\zeta_j,\cdots)^{\cdots 
\varepsilon_{j+1}\varepsilon_j\cdots}
=\sum_{\varepsilon'_j , \varepsilon'_{j+1}=\pm} 
\overline{R}(\zeta_j/\zeta_{j+1})^{
\varepsilon_j , \varepsilon_{j+1}}_{
\varepsilon'_j , \varepsilon'_{j+1}}
\overline{G}_{\sigma}^{(n)} 
(\cdots,\zeta_j,\zeta_{j+1},\cdots)^{
\varepsilon'_j , \varepsilon'_{j+1}}, 
\label{eq:R-symm'} 
\end{equation}
\begin{equation}
\overline{G}_{\sigma}^{(n)} 
(\zeta', x^{2}\zeta_{2n})^{
\varepsilon'\,\varepsilon_{2n}}
=\sigma\overline{G}_{\sigma}^{(n)} 
(\zeta_{2n},\zeta')^{
\varepsilon_{2n}\,\varepsilon'}
\prod_{j=1}^{2n-1} \frac{\zeta_{2n}}{\zeta_j}.
\label{eq:cyc'}
\end{equation}

\subsection{Generalized correlation functions of SOS model}

For later convenience we also 
introduce another meromorphic function 
$Q^{(n)}(w|\zeta )^{kk_1\cdots k_{2n-1}k-2l}$, where 
$0\leqslant l\leqslant n$. 
Since $k_{2n}=k-2l$, the number of elements of 
the following set $A'$ is equal to $n$: 
$$
A'=A \sqcup \{ 2n+1, \cdots , 2n+l \}. 
$$
Let $k_{2n+i}=k_{2n}+2i=k-2(l-i)$ for $1\leqslant i\leqslant l$. 
Then the meromorphic function 
$Q^{(n)}(w|\zeta )^{kk_1\cdots k_{2n-1}k-2l}$ is 
defined as follows: 
\begin{equation}
\begin{array}{rcl}
Q^{(n)}(w|\zeta )^{kk_1\cdots k_{2n-1}k-2l}
&=&\displaystyle\prod_{a\in A} 
\{v_a -u_a +\tfrac{1}{2}-k_a \} \left( 
\prod_{j=1}^{a-1} [v_a -u_j -\tfrac{1}{2}] 
\prod_{j=a+1}^{2n} [u_j -v_a -\tfrac{1}{2}] \right) \\
&\times& (-1)^l \displaystyle\prod_{a'=2n+1}^{2n+l} 
\frac{\{v_{a'}-u_0 +\tfrac{1}{2}-k_{a'} \}}
{[v_{a'}-u_0 -\tfrac{1}{2}]} \left( 
\prod_{j=1}^{2n} [v_{a'} -u_j -\tfrac{1}{2}] \right) \\
&\times& \displaystyle\prod_{j=1}^{2n-1} \{k_j \}^{-1}
\prod_{a,b\in A'\atop a<b} [v_a -v_b +1]^{-1}. 
\end{array}
\label{eq:Q-2}
\end{equation}
When $l=0$, the meromorphic function (\ref{eq:Q-2}) 
is evidently reduced to the expression of (\ref{eq:df-QF}). 

We further introduce the generalized correlation function 
of the cyclic SOS model as follows: 
\begin{equation}
\overline{F}_{\sigma}^{(n)} 
(\zeta)^{kk_1\cdots k_{2n-1}k-2l}=\prod_{a\in A'}\oint_{C_a} 
\dfrac{dw_a}{2\pi\sqrt{-1}w_a} 
\Psi_{\sigma}^{(n)} (w|\zeta )
Q^{(n)}(w|\zeta)^{kk_1\cdots k_{2n-1}k-2l}, 
\label{eq:gG-form}
\end{equation}
where the integral kernel $\Psi_{\sigma}^{(n)} (w|\zeta )$ 
and the integral contour $C_a$ are respectively 
the same ones in the previous section. 

The formula (\ref{eq:gG-form}) can be used only for 
$k_{2n}\leqslant k$. However, by noticing the fact that 
the Boltzmann weights of the cyclic SOS model 
(\ref{eq:HY}) is invariant under the shift $(a,b,c,d)
\mapsto (-a,-b,-c,-d)$, we should obtain the expression 
for $k_{2n}=k+2l>k$ from $\overline{F}_{\sigma}^{(n)} 
(\zeta)^{-k-k_1\cdots -k_{2n-1}-k-2l}$ as follows: 
\begin{equation}
\overline{F}_{\sigma}^{(n)} 
(\zeta)^{kk_1\cdots k_{2n-1}k+2l}=\prod_{a\in A'_-}
\oint_{C_a} \dfrac{dw_a}{2\pi\sqrt{-1}w_a} 
\Psi_{\sigma}^{(n)} (w|\zeta )
Q^{(n)}(w|\zeta)^{kk_1\cdots k_{2n-1}k+2l}, 
\label{eq:gG'-form}
\end{equation}
where 
\begin{equation}
\begin{array}{rcl}
Q^{(n)}(w|\zeta )^{kk_1\cdots k_{2n-1}k+2l}
&=&\displaystyle\prod_{a\in A_-} 
\{v_a -u_a +\tfrac{1}{2}+k_a \} \left( 
\prod_{j=1}^{a-1} [v_a -u_j -\tfrac{1}{2}] 
\prod_{j=a+1}^{2n} [u_j -v_a -\tfrac{1}{2}] \right) \\
&\times& (-1)^l \displaystyle\prod_{a'=2n+1}^{2n+l} 
\frac{\{v_{a'}-u_0 +\tfrac{1}{2}+k_{a'} \}}
{[v_{a'}-u_0 -\tfrac{1}{2}]} \left( 
\prod_{j=1}^{2n} [v_{a'} -u_j -\tfrac{1}{2}] \right) \\
&\times& \displaystyle\prod_{j=1}^{2n-1} \{k_j \}^{-1}
\prod_{a,b\in A'_-\atop a<b} [v_a -v_b +1]^{-1}. 
\end{array}
\label{eq:Q'-2}
\end{equation}
Here 
$$
A_-:=\{ a|k_a =k_{a-1}-1, \,\,1\leqslant a\leqslant 2n \}, 
~~~~ A'_-=A_- \sqcup \{ 2n+1, \cdots , 2n+l \}, 
$$
and 
$k_{2n+i}=k_{2n}-2i=k+2(l-i)$ for $1\leqslant i\leqslant l$. 

\subsection{First integral formula for the XYZ correlation 
functions}

Let 
\begin{equation}
G^{(n)}_\sigma (\zeta_1 , \cdots , \zeta_{2n})=
(\sqrt{-1})^{-n}
\sum_{k_0 , k_1 , \cdots k_{2n}} t^{k_0}_{k_1} (u_1 -u_0 ) 
\otimes \cdots \otimes t^{k_{2n-1}}_{k_{2n}} (u_{2n}-u_0 ) 
F^{(n)}_\sigma (\zeta_1 , \cdots , \zeta_{2n})^{
k_0 k_1 \cdots k_{2n}}. 
\label{eq:G-df}
\end{equation}
Here $\zeta_j =x^{u_j}$, and 
$$
 F_{\sigma}^{(n)}(\zeta)^{kk_1\cdots k_{2n}}
=c_n \prod_{1\leqslant j< k \leqslant 2n} 
\zeta_j^{\frac{r-1}{r}} g(z_j/z_k)
\times \overline{F}_{\sigma}^{(n)}(\zeta)^{kk_1\cdots k_{2n}}, 
$$
where 
$\overline{F}_{\sigma}^{(n)}(\zeta)^{kk_1\cdots k_{2n}}$ 
is defined by (\ref{eq:gG-form},\ref{eq:Q-2}) for 
$k_{2n}\leqslant k$, otherwise by 
(\ref{eq:gG'-form},\ref{eq:Q'-2}). The sum with respect to 
$k_j$ ($1\leqslant j\leqslant 2n$) should be taken over 
$k_j =k_{j-1} \pm 1$. The sum with respect to $k_0$ 
should be as follows. When $r\in \mathbb{Q}$ and 
$2r=N/N'$ ($N$, $N'$ are coprime), we have 
$$
t^{N+k}_{N+k\pm 1}(u)=(-1)^{r+N'} t^{k}_{k\pm 1}(u). 
$$
The cyclic SOS correlation function 
$F^{(n)}_\sigma (\zeta)^{k k_1 \cdots k_{2n}}$ is 
evidently invariant under the shift 
$(k, k_1 , \cdots , k_{2n})\mapsto (k+N, k_1 +N, 
\cdots , k_{2n}+N)$. Thus the summand in (\ref{eq:G-df}) 
is also invariant under the same shift, so that 
the sum with respect to $k_0$ 
can be taken over $\mathbb{Z}/N\mathbb{Z}$. 
Since there is no such invariance, the sum with respect 
to $k_0$ should be taken over $\mathbb{Z}$ when $r$ is 
irrational. From these observations we have 
\begin{equation}
\sum_{k_0} =\left\{ \begin{array}{ll} 
\displaystyle\sum_{k_0 =0}^{N-1} & 
\mbox{if $r\in \mathbb{Q}$, $2r=N/N'$ \; 
($N$, $N'$ are coprime)}; \\
\displaystyle\sum_{k_0\in\mathbb{Z}} & 
\mbox{if $r\not\in \mathbb{Q}$}. 
\end{array} \right. 
\label{eq:sumover}
\end{equation}

Simple observation shows that the $R$-matrix symmetry 
(\ref{eq:R-symm}) follows from the $W$-symmetry 
(\ref{eq:W-symm-comp}) for 
(\ref{eq:gG-form},\ref{eq:gG'-form}). 
The cyclicity for the generalized correlation functions 
in the cyclic SOS model does not hold but we have the 
following relations: 

\begin{prop} For fixed $k_0=k, k_1 , \cdots , k_{2n-2}, 
k_{2n-1}=k'$ with $k_j =k_{j-1}\pm 1$ $(1\leqslant j 
\leqslant 2n-1)$, 
the following cyclicity relations hold: 
\begin{equation}
\begin{array}{cl}
&\displaystyle\sum_{s=\pm 1} 
t^{k'}_{k'+s} (u_{2n}-u_0 +2) 
F_{\sigma}^{(n)} (\zeta_1, \cdots , 
\zeta_{2n-1}, x^{2}\zeta_{2n})^{
kk_1 \cdots k_{2n-2} k' k'+s} \\
=&\displaystyle\sum_{s =\pm 1} 
t^{k-s}_k (u_{2n}-u_0 ) 
F_{\sigma}^{(n)} (\zeta_{2n}, \zeta_1, \cdots , 
\zeta_{2n-1})^{
k-s k k_1 \cdots k_{2n-2} k'}. 
\end{array}
\label{eq:gF-l}
\end{equation}
\label{prop:gF-cyc}
\end{prop}

[Proof] Let $k'=k-2l+1$ with $1\leqslant l \leqslant n$. 
(The case $k'=k+2l-1$ can be similarly proved.) For 
$\varepsilon =\pm$ let 
$$
h_\varepsilon  (u):=
\theta_{2-\varepsilon}\theta_{3+\varepsilon} 
(\tfrac{u}{2r}; \tfrac{\pi\sqrt{-1}}{\epsilon r}) 
=\left\{ \begin{array}{ll} \theta_1 \theta_4 
(\tfrac{u}{2r}; \tfrac{\pi\sqrt{-1}}{\epsilon r}) 
& (\varepsilon >0) \\
\theta_2 \theta_3 (\tfrac{u}{2r}; 
\tfrac{\pi\sqrt{-1}}{\epsilon r}) 
& (\varepsilon <0) \end{array} 
\right. 
$$
Then from (\ref{eq:int-vec},\ref{eq:f-def}) the relation 
(\ref{eq:gF-l}) is reduced to 
\begin{equation}
\begin{array}{cl}
&(-1)^l \mbox{[}u_{2n}-u_0 +1\mbox{]} 
\displaystyle\sum_{s =\pm 1} 
h_\varepsilon  (k'-s(u_{2n}-u_0 +2)) 
F_{\sigma}^{(n)} (\zeta', x^{2}\zeta_{2n})^{
kk_1 \cdots k_{2n-2} k' k'+s} \\
+&\mbox{[}u_{2n}-u_0+2\mbox{]} 
\displaystyle\sum_{s =\pm 1} s h_\varepsilon 
(k-s(u_{2n}-u_0+1)) 
F_{\sigma}^{(n)} (\zeta_{2n}, \zeta')^{
k-s k k_1 \cdots k_{2n-2} k'}=0. 
\end{array}
\label{eq:equivgF-l}
\end{equation}
Here we used the abbreviation $\zeta' =
(\zeta_1 , \cdots , \zeta_{2n-1})$ and the 
functional relation 
$$
\frac{f(u+2)}{f(u)}=\frac{[u+1]}{[u+2]}. 
$$
Simple calculations show that 
\begin{equation}
\begin{array}{cl}
& \displaystyle\sum_{s =\pm 1} 
h_\varepsilon  (k'-s(u_{2n}-u_0)) 
F_{\sigma}^{(n)} (\zeta', \zeta_{2n})^{
kk_1 \cdots k_{2n-2} k' k'+s} \\
=&\displaystyle\prod_{a\in A'}\oint_{C_a} 
\dfrac{dw_a}{2\pi\sqrt{-1}w_a} 
\Psi_{\sigma}^{(n)} (w_{a_1} \cdots , 
w_{2n+l}|\zeta_1 , \cdots \zeta_{2n} )
{\cal F}_{\varepsilon,l} (v|u_0 , u_1 , \cdots , u_{2n}), 
\end{array}
\label{eq:2n-sum}
\end{equation}
where 
\begin{equation}
\begin{array}{cl}
&{\cal F}_{\varepsilon,l} (v|u_0 , u_1 , \cdots u_{2n}) =
h_\varepsilon (k-2l+2+u_0 +u_{2n}-2v_{2n+1}) \\
\times& \displaystyle\frac{[u_{2n}-u_0 ]}
{[v_{2n+1}-u_0 -\frac{1}{2}]}
\prod_{i=2}^l \frac{\{v_{2n+i}-u_0 +\tfrac{1}{2}-k
+2(l-i) \}}
{[v_{2n+i}-u_0 -\tfrac{1}{2}]} 
[ v_{2n+i}-u_{2n} -\tfrac{1}{2} ] \\
\times& \displaystyle\prod_{a\in A\atop a<2n} 
\{v_a -u_a +\tfrac{1}{2}-k_a \} \left( 
\prod_{j=1}^{a-1} [v_a -u_j -\tfrac{1}{2}] 
\prod_{j=a+1}^{2n} [u_j -v_a -\tfrac{1}{2}] \right) \\
\times& (-1)^{l-1} \displaystyle\prod_{a'=2n+1}^{2n+l} 
\prod_{j=1}^{2n-1} [v_{a'} -u_j -\tfrac{1}{2}] 
\prod_{j=1}^{2n-2} \{k_j \}^{-1}
\prod_{a,b\in A'\atop a<b} [v_a -v_b +1]^{-1}. 
\end{array}
\label{eq:exp-Fl}
\end{equation}
It follows from the definition that $2n\in A$ and 
$A'=A \sqcup \{ 2n+1, \cdots , 2n+l-1 \}$ when $s=1$, 
and that $2n\not\in A$ and 
$A'=A \sqcup \{ 2n+1, \cdots , 2n+l \}$ when $s=-1$ 
in the LHS of (\ref{eq:2n-sum}). Thus 
we made the shift of 
variables $(w_{2n}, \cdots , w_{2n+l-1})\mapsto 
(w_{2n+1}, \cdots , w_{2n+l})$ for $s=1$ in 
(\ref{eq:2n-sum}). 
Furthermore, in order to derive the expression of 
(\ref{eq:exp-Fl}) we used the identity 
$$
\begin{array}{cl}
&\dfrac{[v-u-\tfrac{1}{2}]}{[v-u_0 -\tfrac{1}{2}]}
\{ v-u_0 -\tfrac{1}{2}-k \} 
h_\varepsilon (k+u-u_0 ) -
\{ v-u -\tfrac{1}{2}-k \} 
h_\varepsilon (k-u+u_0 )\\
=&-\dfrac{[u-u_0 ]\{k\}
h_\varepsilon (k+1+u_0 +u -2v)}{[v-u_0 -\tfrac{1}{2}]}. 
\end{array}
$$
By repeating the similar calculations we find that 
eq. (\ref{eq:equivgF-l}) is reduced to 
\begin{equation}
\begin{array}{cl}
\displaystyle\prod_{a\in A'}\oint_{C_a} 
\dfrac{dw_a}{2\pi\sqrt{-1}w_a} 
\Psi_{\sigma}^{(n)} (w|\zeta_1 , \cdots \zeta_{2n} )
{\cal A}_l (v|u)
{\cal B}_{\varepsilon,k,l}
(v_{2n+1}, \cdots , v_{2n+l}|u_0 , u_{2n})=0, 
\end{array}
\label{eq:rec-Fl}
\end{equation}
where 
$$
\begin{array}{rcl}
{\cal A}_l &=& \displaystyle\prod_{a\in A\atop a<2n} 
\{v_a -u_a +\tfrac{1}{2}-k_a \} \left( 
[ v_a -u_{2n} -\tfrac{1}{2}] 
\prod_{j=1}^{a-1} [v_a -u_j -\tfrac{1}{2}] 
\prod_{j=a+1}^{2n-1} [u_j -v_a -\tfrac{1}{2}] \right) \\
&\times& \displaystyle\prod_{i=1}^{2n+l} 
[v_{2n+i} -u_0 +\tfrac{3}{2}]^{-1} 
\left( \prod_{j=1}^{2n-1} [u_j -v_{2n+i} -\tfrac{1}{2}] 
\right) \\
&\times&\displaystyle\prod_{j=1}^{2n-2} \{k_j \}^{-1}
\prod_{a,b\in A'\atop a<b<2n} [v_a -v_b +1]^{-1}
\prod_{a,a'\in A'\atop a<2n<a'} [v_a -v_{a'}-1]^{-1}
\prod_{a',b'\in A'\atop 2n<a'<b'} [v_{a'}-v_{b'}+1]^{-1}, \\
{\cal B}_{\varepsilon,k,l} &=& 
h_\varepsilon (k-2l +u_0 +u_{2n}-2v_{2n+1}) 
\displaystyle\prod_{i=2}^l \{ v_{2n+i} -u_0 +
\tfrac{5}{2}-k+2(l-i)\} [v_{2n+i}-u_{2n}-\tfrac{1}{2}] \\
&+& (-1)^l h_\varepsilon (k-2+u_0 +u_{2n}-2v_{2n+l}) 
\displaystyle\prod_{i=1}^{l-1} \{ v_{2n+i} -u_0 +
\tfrac{3}{2}-k+2(l-i)\} [u_{2n}-v_{2n+i}-\tfrac{1}{2}]. 
\end{array}
$$

It is evident that (\ref{eq:rec-Fl}) for $l=1$ holds because 
${\cal B}_{\varepsilon,k,1}=0$. In order to prove (\ref{eq:rec-Fl}) 
for $l>1$, we wish to show that 
\begin{equation}
{\cal B}_{\varepsilon,k,l}(v|u_0 , u_{2n})=
\sum_{i=1}^{l-1} [v_{2n+i}-v_{2n+i+1}+1] 
B_{\varepsilon,k,l}^{(i)} (v|u_0 , u_{2n}), 
\label{eq:id-Bl}
\end{equation}
where $B_{\varepsilon,k,l}^{(i)}$ is some function that is 
symmetric with respect to $v_{2n+i}$ and $v_{2n+i+1}$. 
Suppose that (\ref{eq:id-Bl}) is true. Then 
${\cal A}_l {\cal B}_{\varepsilon,k,l}$ is 
the sum of $l-1$ terms, each of which is symmetric 
with respect to $v_{2n+i}$ and $v_{2n+i+1}$. Correspondingly, 
the integral (\ref{eq:rec-Fl}) vanishes from the antisymmetry 
of $\Psi_{\sigma}^{(n)}$ with respect to $w$'s. 
The claim of this Proposition is therefore reduced to 
eq. (\ref{eq:id-Bl}). 

Let us show (\ref{eq:id-Bl}) by the induction for $l>1$. 
The validity of eq. (\ref{eq:id-Bl}) for $l=2$ 
follows from the following identity: 
\begin{equation}
\begin{array}{cl}
&{\cal B}_{\varepsilon,k,2} (v_{2n+1}, v_{2n+2}|u_0 , u_{2n})\\
=& h_\varepsilon (k-4 +u_0 +u_{2n}-2v_{2n+1}) 
\{ v_{2n+2}-u_0 +\tfrac{5}{2}-k \} 
[v_{2n+2}-u_{2n}-\tfrac{1}{2}] \\
+& h_\varepsilon (k-2 +u_0 +u_{2n}-2v_{2n+2}) 
\{ v_{2n+1}-u_0 +\tfrac{7}{2}-k \} 
[u_{2n}-v_{2n+1}-\tfrac{1}{2}] \\
=& -h_\varepsilon (k-3 +u_0 -u_{2n}) 
\{ k-3+u_0 +u_{2n} -v_{2n+1} -v_{2n+2} \} 
\mbox{[}v_{2n+1}-v_{2n+2}+1\mbox{]}. 
\end{array}
\label{eq:B2}
\end{equation}

For $l>2$, the assumption of the induction implies 
$$
{\cal B}_{\varepsilon,k-2,l-1}
(v_{2n+1}, \cdots , v_{2n+l-1}|u_0 , 
u_{2n})=\sum_{i=1}^{l-2} [v_{2n+i}-v_{2n+i+1}+1] 
B_{\varepsilon,k-2,l-1}^{(i)} 
(v_{2n+1}, \cdots , v_{2n+l-1}|u_0 , u_{2n}). 
$$
The validity of eq. (\ref{eq:id-Bl}) for general $l$ 
thus follows from the relation 
$$
\begin{array}{cl}
&{\cal B}_{\varepsilon,k,l}
(v_{2n+1}, \cdots , v_{2n+l}|u_0 , u_{2n}) \\
=&{\cal B}_{\varepsilon,k-2,l-1} (v_{2n+1}, \cdots , v_{2n+l-1}
|u_0 , u_{2n})\{ v_{2n+l}-u_0 +\tfrac{5}{2}-k\} 
[v_{2n+l}-u_{2n}-\tfrac{1}{2}] \\
+&(-1)^l \displaystyle{\cal B}_{\varepsilon,k,2} 
(v_{2n+l-1}, v_{2n+l}|u_0 , u_{2n})\prod_{i=1}^{l-2} 
\{ v_{2n+i}-u_0 +\tfrac{3}{2}-k+2(l-i)\} 
[u_{2n}-v_{2n+i}-\tfrac{1}{2}], 
\end{array}
$$
where we again use (\ref{eq:B2}). $\Box$ 

As a Corollary of Proposition \ref{prop:gF-cyc}, 
we have 
\begin{cor} The cyclicity (\ref{eq:cyc}) holds for the 
XYZ correlation functions (\ref{eq:G-df}). 
\label{cor:G-cyc}
\end{cor}

\subsection{Normalization in the XYZ case}

Furthermore, we find that the following 
normalization condition holds for the generalized 
correlation functions in the cyclic SOS model for 
$k_{2n}\equiv k$ (mod $2$): 
\begin{equation}
\begin{array}{rcl}
F_{\sigma}^{(n)} (\zeta'', 
\zeta_{2n-1}, sx^{-1}\zeta_{2n-1})^{
k\cdots k' k'\pm 1 k_{2n}} &=& \delta_{k',k_{2n}}
\frac{d_s}{\{k_{2n}\}} 
F_{\sigma}^{(n-1)} 
(\zeta'')^{k\cdots k'} 
\;\;\;\; (s=\pm ), 
\end{array}
\label{eq:rec-gF-comp}
\end{equation}
where 
$$
d_s =\left\{ \begin{array}{ll} 
1 & (s>0); \\
e^{-\frac{\pi\sqrt{-1}}{2r}(1+r \pm 2k)} 
& (s<0). \end{array} \right. 
$$

\begin{prop}
The normalization condition (\ref{eq:rec-G}) holds 
for the XYZ correlation functions (\ref{eq:G-df}). 
\label{prop:Normal}
\end{prop}

[Proof] First we notice the following identity 
\begin{equation}
\begin{array}{cl}
& \theta_1\theta_4 (\tfrac{k+u}{2r}; 
\tfrac{\pi\sqrt{-1}}{\epsilon r}) 
\theta_2\theta_3 (\tfrac{k-u}{2r}; 
\tfrac{\pi\sqrt{-1}}{\epsilon r}) -
\theta_1\theta_4 (\tfrac{k-u}{2r}; 
\tfrac{\pi\sqrt{-1}}{\epsilon r}) 
\theta_2\theta_3 (\tfrac{k+u}{2r}; 
\tfrac{\pi\sqrt{-1}}{\epsilon r}) \\
=&
\theta_2\theta_3 (0; 
\tfrac{\pi\sqrt{-1}}{\epsilon r})
\theta_1 (\tfrac{u}{2r}; 
\tfrac{\pi\sqrt{-1}}{\epsilon r})
\theta_4 (\tfrac{k}{2r}; 
\tfrac{\pi\sqrt{-1}}{\epsilon r})=C^2 [u]\{ k\}, 
\end{array}
\label{eq:ad-thm}
\end{equation}
where $C$ is the same constant as the one defined in 
(\ref{eq:C-df}). Using (\ref{eq:ad-thm}) and 
(\ref{eq:C-df}), we have 
\begin{equation}
\begin{array}{cl}
& \left. \displaystyle\sum_{s=\pm 1} 
t^k_{k+s}(u_{2n-1}-u_0 )^{\varepsilon_{2n-1}}
t^{k+s}_k (u_{2n}-u_0 )^{\varepsilon_{2n}}\right|_{
u_{2n}=u_{2n-1}-1} \\
=& \sqrt{-1} \{ k\} \delta_{\varepsilon_{2n-1}+
\varepsilon_{2n},0}. 
\end{array}
\label{eq:t-ad}
\end{equation}
Furthermore, we also have 
\begin{equation}
\begin{array}{cl}
& \left. \displaystyle\sum_{s=\pm 1} 
e^{-s\frac{\pi\sqrt{-1}k}{r}} 
t^k_{k+s}(u_{2n-1}-u_0 )^{\varepsilon_{2n-1}}
t^{k+s}_k (u_{2n}-u_0 )^{\varepsilon_{2n}}\right|_{
u_{2n}=u_{2n-1}-1-\frac{\pi\sqrt{-1}}{\epsilon}} \\
=& \varepsilon_{2n}e^{-\frac{\pi\sqrt{-1}}{2r}}  \{ k\} 
\delta_{\varepsilon_{2n-1}+
\varepsilon_{2n},0}. 
\end{array}
\label{eq:tm-ad}
\end{equation}
Thus, the normalization condition (\ref{eq:rec-G}) in 
the XYZ Heisenberg antiferromagnet follows from 
(\ref{eq:t-ad}), (\ref{eq:tm-ad}) 
and (\ref{eq:rec-gF-comp}). $\Box$

{}From Corollary \ref{prop:gF-cyc} 
and Proposition \ref{prop:Normal}, 
we have the following Theorem. 

\begin{thm} The XYZ correlation function 
$G^{(n)}_\sigma (\zeta )$ (\ref{eq:G-df}) 
solves the bootstrap equations 
(\ref{eq:R-symm}--\ref{eq:rec-G}). 
\label{thm:LaP}
\end{thm}

\subsection{Second integral formula for the XYZ correlation 
functions}

In this subsection we assume that $r\in \mathbb{Z}_{>1}$. 
The integral formula (\ref{eq:G-df}) remains the sum 
with respect to the local height variables $k, k_1 , 
\cdots , k_{2n}$ in the cyclic SOS model. Thus, the 
expression (\ref{eq:G-df}) consists of $2r\times 2^{2n}$ 
terms because of (\ref{eq:sumover}) and the condition 
$k_j =k_{j-1}\pm 1$ for $1\leqslant j\leqslant 2n$. 

Now we wish to present another integral solution 
to the bootstrap equations (\ref{eq:R-symm}--\ref{eq:rec-G}). 
Eq. (\ref{eq:2n-sum}) shows that the sum with 
respect to $k_{2n}$ with fixed $k, k_1 , \cdots , k_{2n-1}$ 
can be performed to be in a simple form. In order to repeat 
this procedure with respect to $k_{2n-1}, \cdots k_1$, 
we find that the number of integral variables should be 
greater than $2n$ for the $2n$-point correlation function. 

Let $k_{2n}=k+2n-2l$ ($0\leqslant l\leqslant 2n$). 
Then the number of the set $A$ is equal to $2n-l$, 
where 
$$
A:=\{ a| k_a =k_{a-1}+1, \,\,1\leqslant a\leqslant 2n \}. 
$$
We further introduce the set of indices 
$$
A''=A \sqcup \{ 2n+1, \cdots , (r+1)n+l \}. 
$$
Note that the number of the set $A''$ is equal to $(r+1)n$. 
Let $k_{2n+i}=k_{2n}+2i=k+2(n-l+i)$ for 
$1\leqslant i\leqslant (r-1)n+l$. Then the meromorphic function 
$Q^{(n)}(w|\zeta )^{kk_1\cdots k_{2n}}$ with $k_{2n}=
k+2n-2l$ is defined as follows: 
\begin{equation}
\begin{array}{rcl}
Q^{(n)}(w|\zeta )^{kk_1\cdots k_{2n}}
&=&\displaystyle\prod_{a\in A} 
\{v_a -u_a +\tfrac{1}{2}-k_a \} \left( 
\prod_{j=1}^{a-1} [v_a -u_j -\tfrac{1}{2}] 
\prod_{j=a+1}^{2n} [u_j -v_a -\tfrac{1}{2}] \right) \\
&\times& (-1)^{n-l} \displaystyle\prod_{a'=2n+1}^{(r+1)n+l} 
\frac{\{v_{a'}-u_0 +\tfrac{1}{2}-k_{a'} \}}
{[v_{a'}-u_0 -\tfrac{1}{2}]} \left( 
\prod_{j=1}^{2n} [v_{a'} -u_j -\tfrac{1}{2}] \right) \\
&\times& \displaystyle\prod_{j=1}^{2n-1} \{k_j \}^{-1}
\prod_{a,b\in A''\atop a<b} [v_a -v_b +1]^{-1}. 
\end{array}
\label{eq:Q''-2}
\end{equation}
Set 
\begin{equation}
\tilde{F}_{\sigma}^{(n)}(\zeta)^{kk_1\cdots k_{2n}}
=\tilde{c}_n \prod_{1\leqslant j< k \leqslant 2n} 
\zeta_j^{\frac{r-1}{r}} g(z_j/z_k)
\times \overline{\tilde{F}}_{\sigma}^{(n)}
(\zeta)^{kk_1\cdots k_{2n}}, 
\label{eq:genF-bar}
\end{equation}
where $\tilde{c}_n$ is some constant, and 
\begin{equation}
\overline{\tilde{F}}_{\sigma}^{(n)} 
(\zeta)^{kk_1\cdots k_{2n}}=\prod_{a\in A''}\oint_{C_a} 
\dfrac{dw_a}{2\pi\sqrt{-1} w_a} 
\tilde{\Psi}_{\sigma}^{(n)} (w|\zeta )
Q^{(n)}(w|\zeta)^{kk_1\cdots k_{2n}}. 
\label{eq:genF-form}
\end{equation}
Here, the kernel has the form 
\begin{equation}
\tilde{\Psi}^{(n)}_\sigma (w| \zeta )=
\tilde{\vartheta}^{(n)}_\sigma (w | \zeta )
\prod_{a''\in A}\prod_{j=1}^{2n} x^{-\frac{(v_a -u_j)^2}{2r}} 
\psi \Bigl(\frac{w_{a}}{z_j}\Bigr) 
\prod_{1\leqslant j<k\leqslant 2n} 
x^{-\frac{(u_j -u_k)^2}{4r}}. 
\label{eq:df-tPsi}
\end{equation}
The function 
$\tilde{\vartheta}^{(n)}_\sigma(w|\zeta )$ is a function 
of $(r+1)n$ $w$'s and $2n$ $\zeta$'s. The properties of 
$\tilde{\vartheta}^{(n)}_\sigma(w|\zeta )$ is the same 
as those of $\vartheta^{(n)}_\sigma(w|\zeta )$, but 
in the RHS of (\ref{eq:z-sym}) the product with 
respect to $a$ should be taken over $A''$. 

The second integral formula of the XYZ correlation 
functions is as follows: 
\begin{equation}
\tilde{G}^{(n)}_\sigma (\zeta)=
(\sqrt{-1})^{-n}
\sum_{k_0 , k_1 , \cdots k_{2n}} t^{k_0}_{k_1} (u_1 -u_0 ) 
\otimes \cdots \otimes t^{k_{2n-1}}_{k_{2n}} (u_{2n}-u_0 ) 
\tilde{F}^{(n)}_\sigma (\zeta)^{k_0 k_1 \cdots k_{2n}}. 
\label{eq:genG-df}
\end{equation}
This formula essentially solves the first two of the 
bootstrap equations. The $R$-matrix symmetry 
(\ref{eq:R-symm}) evidently holds. The modified cyclicty 
\begin{equation}
P_{12}\cdots P_{2n-1 2n} 
\tilde{G}_{\sigma}^{(n)} (\zeta', x^{2}\zeta_{2n})
=\sigma (-1)^{(r-1)n}
\tilde{G}_{\sigma}^{(n)} (\zeta_{2n}, \zeta'), 
\label{eq:mod-cyc}
\end{equation}
can be proved in a similar manner as in Proposition 
\ref{prop:gF-cyc} and Corollary \ref{cor:G-cyc}. 

The advantage of the second formula consists in the 
fact that the sum with respect to $k_1 , \cdots , k_{2n}$ 
can be carried out. By repeating the similar calculation 
as (\ref{eq:2n-sum}) was derived, we find 
$$
\tilde{G}^{(n)}_\sigma (\zeta)=\prod_{a\in A''}\oint_{C_a} 
\dfrac{dw_a}{2\pi\sqrt{-1}w_a} 
\tilde{\Psi}_{\sigma}^{(n)} (w|\zeta )
\tilde{Q}^{(n)}(w|\zeta), 
$$
where 
\begin{equation}
\begin{array}{rcl}
\tilde{Q}^{(n)}(w|\zeta)^{\varepsilon_1 \cdots 
\varepsilon_{2n}}&=&\displaystyle\sum_{k=0}^{2r-1} 
(\sqrt{-1})^{-n+\sum_{j=0}^{2n}\varepsilon_j (k-2n-1+j)} 
(-1)^{(r+1)n} \\
&\times & \displaystyle\prod_{a=1}^{2n} 
\frac{[u_a -u_0 ]}{[v_a -u_0 -\tfrac{1}{2}]} 
\left( \prod_{j=1}^{a-1} [v_a -u_j -\tfrac{1}{2}] 
\prod_{j=a+1}^{2n} [u_j -v_a -\tfrac{1}{2}]\right) \\
&\times & \displaystyle\prod_{i=1}^{(r-1)n} 
\frac{\{v_{2n+i} -u_0 +\tfrac{1}{2}-k-2(n+i)\}}
{[v_{2n+i} -u_0 -\tfrac{1}{2}]} \left( 
\prod_{j=1}^{2n} [v_{2n+i}-u_j -\tfrac{1}{2}] \right) \\
&\times & \displaystyle\prod_{j=1}^{2n} f(u_j -u_0 )
h_{\varepsilon_j} (k+j+u_0 +u_j -2v_j ) 
\prod_{a,b\in A''\atop a<b} [v_a -v_b +1]^{-1}. 
\end{array}
\end{equation}
Finally, we have to carry out the sum with respect to $k$ 
of the following form: 
\begin{equation}
\begin{array}{rcl}
F^{(n)}(u_0 |\{ u_j \}_{1\leqslant j\leqslant 2n}|
\{ v_a \}_{1\leqslant a\leqslant (r+1)n})^\varepsilon
&=&\displaystyle\sum_{k=0}^{2r-1} 
(\sqrt{-1})^{k\sum_{j=0}^{2n}\varepsilon_j} 
\displaystyle\prod_{j=1}^{2n} 
h_{\varepsilon_j} (k+j+u_0 +u_j -2v_j ) \\
&\times&\displaystyle\prod_{i=1}^{(r-1)n} 
\{v_{2n+i} -u_0 +\tfrac{1}{2}-k-2(n+i)\}. 
\end{array}
\end{equation}
This function $F^{(n)}$ has precise quasi-periodicities 
as a function of $u_0$, $u_j$ ($1\leqslant j\leqslant 2n$) and 
$v_a$ ($1\leqslant a\leqslant (r+1)n$). 
In particular, you can easily find the 
$u_0$-dependent part of $F^{(1)}$. Furthermore, 
if the limit $r\rightarrow 1$ is taken, 
the explicit expression for $n=1$ can be obtained 
as follows: 
\begin{equation}
\begin{array}{cl}
& F^{(1)}(u_0 |u_1 , u_2 |v_1 , v_2 )^{\varepsilon_1 
\varepsilon_2} \\
=&\left\{ 
\begin{array}{ll} C_- \theta_2 \left( \tfrac{u_1 -u_2}{2}
-v_1 +v_2 ; \tfrac{\pi\sqrt{-1}}{\epsilon} \right) 
\theta_3 \left( u_0 +\tfrac{u_1 +u_2}{2}
-v_1 -v_2 ; \tfrac{\pi\sqrt{-1}}{\epsilon} \right) 
& (\varepsilon_1 \varepsilon_2 <0); \\ 
C_+ \theta_1 \left( \tfrac{u_1 -u_2}{2}
-v_1 +v_2 ; \tfrac{\pi\sqrt{-1}}{\epsilon} \right) 
\theta_4 \left( u_0 +\tfrac{u_1 +u_2}{2}
-v_1 -v_2 ; \tfrac{\pi\sqrt{-1}}{\epsilon} \right) 
& (\varepsilon_1 \varepsilon_2 >0). 
\end{array} \right. 
\end{array}
\end{equation}
Here $C_\pm$ are some constants times exponential 
functions of $u_0, u_1, u_2 , v_1, v_2$. 

\section{Concluding remarks} 

In this paper we have constructed two integral 
solutions to the bootstrap equations for the 
XYZ Heisenberg antiferromagnet. These solutions 
are expected to give the correlation functions of 
the XYZ model in the antiferromagnetic regime. 
The first solution is essentially the same as the 
formula given by Lashkevich and Pugai \cite{LaP}. 
Lashkevich-Pugai's formula can be obtained from 
the correlation function in the RSOS type model, 
by using the vertex-face correspondence. In order to 
avoid the pole resulting from $[k]$ in the denominator, 
$k\in \mathbb{Z}+\delta$ should be assumed with 
some real $\delta$, and the limit $\delta\rightarrow 0$ 
should be taken after all calculation \cite{LaP}. On the 
other hand, we constructed our first formula from 
the correlation function in the cyclic SOS model 
so that no such regularization was needed. 

By the construction, both Lashkevich-Pugai's formula 
and our first formula for $2n$-point XYZ correlation 
function remain the sum with respect to 
the local height variables $k_0 , k_1 , \cdots , k_{2n}$ 
other than $n$-fold integral. 
When $r\in\mathbb{Z}_{>1}$ we construct another 
$(r+1)n$-fold integral formula, in which the sum with 
respect to $k_1 , \cdots , k_{2n}$ can be carried out. 
The structure of our second formula 
is quite similar to Shiraishi's formula \cite{Shi}, though 
the latter treated only the case $r=\frac{3}{2}$. 

Concerning the second formula, we have not yet 
proved the normalization condition. The number 
of integral variables in the second formula are much 
greater than that in the first one, so that the recursion 
relation for the second one is not so simple. Nevertheless, 
we believe that the normalization condition holds in the 
second integral formula. Actually, the number of integral 
variables in the second formula is not minimal but 
redundant. It is sufficient for the number of integral 
variables $N$ in the second formula to 
satisfy $n \leqslant N-n \equiv 0$ (mod $r$). Thus, we 
can make an $N(n)$-fold integral formula, where 
$N(n)=n+mr$ for 
$(m-1)r<n\leqslant mr$ with some $m\in\mathbb{Z}_{>0}$. 
If we employ this `minimal' integral solution, 
the normalization condition holds unless $n\equiv 1$ 
(mod $r$) from the same reason as in the first formula. 
However, the validity of the recursion relation 
in the `minimal' integral solution becomes again unclear 
for $n=mr+1$ because $N(n)-N(n-1)=r+1$. 

Another future problem is to consider the bootstrap equations 
for Belavin's $\mathbb{Z}/n\mathbb{Z}$-symmetric model and 
to construct integral formulae of the correlation functions. 
We wish to address this problem in a separate paper. 

\section*{Acknowledgements}
I would like to thank M. Lashkevich, A. Nakayashiki and 
J. Shiraishi for valuable discussion. I would also 
like to dedicate the present paper to the memory of 
Sung-Kil Yang, who had been interested in 
{\it Infinite Analysis} projects including my studies. 

\appendix

\section{Properties of the $R$-matrix}

The main properties of the $R$-matrix of the eight-vertex 
model are the Yang-Baxter equation
\begin{equation}
R_{12}(\zeta_1/\zeta_2)
R_{13}(\zeta_1/\zeta_3)
R_{23}(\zeta_2/\zeta_3)=
R_{23}(\zeta_2/\zeta_3)
R_{13}(\zeta_1/\zeta_3)
R_{12}(\zeta_1/\zeta_2); 
\label{eq:YBE}
\end{equation}
where the subscript of the $R$-matrix denotes 
the spaces on which $R$ nontrivially acts; 
the initial condition 
\begin{equation}
R(1)=P; 
\label{eq:ini}
\end{equation}
the unitarity relation
\begin{equation}
R_{12}(\zeta_1/\zeta_2)R_{21}(\zeta_2/\zeta_1)=1; 
\label{eq:uni}
\end{equation}
the $\mathbb{Z}_2$-parity
\begin{equation}
R_{12}(-\zeta)=-\sigma^z_1 R_{12}(\zeta)\sigma^z_1 ; 
\label{eq:parity}
\end{equation}
and the crossing symmetries
\begin{equation}
R_{21}^{t_1}(\zeta_2/\zeta_1)=\sigma^x_1 
    R_{12}(x^{-1}\zeta_1/\zeta_2)\sigma^x_1 
    =-\sigma^y_1 
    R_{12}(-x^{-1}\zeta_1/\zeta_2)\sigma^y_1. 
\label{eq:cross}
\end{equation}
In (\ref{eq:parity},\ref{eq:cross}) the shift 
$\zeta\mapsto -\zeta$ implies the one such that 
$u\mapsto u-\frac{\pi\sqrt{-1}}{\epsilon}$. 
The properties (\ref{eq:ini}--\ref{eq:cross}) hold 
if the normalization factor of the $R$-matrix satisfies 
the following relations: 
\begin{equation}
\kappa (\zeta )\kappa (\zeta^{-1})=1, ~~~~
\kappa (x^{-1}\zeta^{-1})=\kappa (\zeta). 
\label{eq:UC}
\end{equation}
Under this normalization the partition function per 
lattice site is equal to unity in the thermodynamic 
limit \cite{ESM,JMbk}.

\end{document}